\documentstyle[version2,aps]{revtex}
\draft
\begin{document}
\columnsep -.375in
\twocolumn[
\begin{title}
Toulouse Points and Non-Fermi Liquid States \\
in the Mixed Valence Regime of the Generalized Anderson Model
\end{title}

\author{Gabriel Kotliar}

\begin{instit}
Serin Physics Laboratory, Rutgers University,
Piscataway, NJ 08855-0849, USA
\end{instit}

\author{Qimiao Si\cite{rice}}
\begin{instit}
Department of Physics, University of Illinois,
Urbana, IL 61801, USA
\end{instit}

\begin{instit}
\end{instit} 
 
\begin{abstract}
We study the mixed valence regime of a generalized Anderson impurity
model using the bosonization approach. This single impurity problem
is defined by the $U=\infty$ Anderson model with an additional
density-density interaction, as well as an explicit exchange
interaction, between the impurity and conduction electrons.
We find three points in the interaction parameter space at which
all the correlation functions can be calculated explicitly.
These points represent the mixed valence counterparts of the
usual Toulouse point for the Kondo problem, and are appropriately
named the Toulouse points of the mixed valence problem.
Two of the Toulouse points exhibit the strong coupling, Fermi
liquid behavior. The third one shows spin-charge separation;
here, the spin-spin correlation functions are Fermi-liquid-like,
the charge-charge correlation functions and the single particle
Green function have non-Fermi-liquid behaviors, and a pairing
correlation function is enhanced compared to the Fermi
liquid case. This third Toulouse point describes the novel
intermediate mixed valence phase we have previously identified.
In deriving these results, we emphasize the importance of keeping
track of the anticommutation relation between the fermion fields
when the bosonization method is applied to quantum impurity problems.
\end{abstract}

\vskip 0.5 in
]
\narrowtext
\section{\bf Introduction}

\label{sec:intro}

The mixed valence problem is a classic problem in condensed matter
theory. It describes an impurity  with three configurations,
having both charge and spin degrees of freedom, coupled to a 
conduction electron bath. It differs from the Kondo problem 
in that low lying  local  charge fluctuations coexist with local  
spin fluctuations. The Kondo problem has been studied by a variety
of techniques, and is by now well understood. As long as
the effective Kondo coupling is antiferromagnetic, the low
energy behavior is described by a strong coupling, Fermi liquid
fixed point\cite{AndersonYuval,Wilson,Nozieres,AndreiWiegmann}. 
The mixed valence problem has also been studied extensively.
A variational study by Varma and Yafet\cite{VarmaYafet}, and 
renormalization group studies of Haldane\cite{Haldane} and 
Krishnamurthy {\it et al.}\cite{Krishnamurthy}, have all found
that the low energy behavior of the mixed valence problem is
described by a strong coupling, Fermi liquid fixed point.
This fixed point is qualitatively similar to that of the Kondo 
problem, though quantities such as the Wilson ratio are modified.
These works were followed by extensive studies on the proper
description of the resulting Fermi liquid states of both
the Anderson impurity and Anderson lattice problems.
These later works used techniques ranging from
Gutzwiller variational wavefunctions to the slave boson large
N expansion\cite{Bickers}. They have provided qualitative,
and sometimes quantitative, understandings of the physical properties
of the heavy fermion metals.

In a series of papers we have revisited the mixed valence problem 
in an attempt to identify metallic non-Fermi liquid phases in a 
two band extended Hubbard model\cite{SiKotliar1,SiKotliar2,KotliarSi}.
In the limit of infinite dimensions, this extended Hubbard model
can be solved through a generalized Anderson impurity model with a 
self-consistent electron bath. This single impurity model is 
defined by the $U=\infty$ Anderson model with an additional 
density-density interaction, as well as an explicit exchange 
interaction, between the impurity and conduction electrons. 
Away from half-filling, the self-consistency condition 
implies that, the associated impurity problem is in the mixed 
valence regime over an extended range of densities. The persistence
of the mixed valence behavior for the self-consistent impurity 
problem associated with a lattice model in infinite dimensions
is in fact quite general and is not restricted to the two 
band extended Hubbard model. For example, in the one band case, 
the average occupation of the impurity in the self-consistent 
Anderson model is simply the lattice density, which is not 
very close to one (local moment regime) or zero (empty orbital 
regime) over a wide range of densities. This, 
together with our finding that away from half-filling 
the self-consistent conduction electron bath has a finite density
of states at the Fermi level, imply that the classification of 
the fixed points in the mixed valence regime of the generalized 
Anderson impurity model with a regular electron bath also 
specifies the possible metallic phases of the extended Hubbard 
model in infinite dimensions. Of course, in general, the couplings 
of the effective Anderson model at the cutoff energy scale of the 
universal regime can be different from the bare atomic interactions, 
just like the Coulomb pseudopotential at the Debye temperature 
scale is different from the bare Coulomb 
interactions\cite{MorelAnderson}.

We studied the generalized Anderson impurity model by extending
Haldane's renormalization group scheme such that the local charge
fluctuations and local spin fluctuations are treated on an equal
footing. In the mixed valence regime, we found three, and only
three, kinds of fixed points, which we termed the strong coupling,
weak coupling, and intermediate phase, respectively.
The strong coupling and weak coupling phases are the direct
analogs of their counterparts in the Kondo problem.
The intermediate phase is entirely new and occurs only in the
mixed valence regime. Its existence came as a surprise.
In this new phase, spin and charge excitations are separated.
From the renormalization group analysis, it is expected that
the spin-spin correlation functions remain to have the Fermi
liquid form, while the charge-charge correlation functions and
single particle Green function have an algebraic
behavior with interaction-dependent exponents.

The renormalization group procedure is based on a Coulomb gas
representation for the single impurity problem. As usual, the
Coulomb gas analysis uses a dilute instanton expansion. In the
strong coupling and intermediate phases, some of the fugacities
flow towards strong coupling, and the behaviors of the correlation
functions can only be inferred through ``extrapolating'' the
scaling trajectories beyond the dilute instanton regime. For the
intermediate phase, this procedure
does not allow an explicit determination of the exponents. Finally,
it is in principle possible that additional fixed points, not
captured by the dilute instanton expansion, may occur. An example for
the latter arises in the related, though qualitatively different,
problem of tunneling through a point contact in a Luttinger
liquid\cite{KaneFisher,Furusaki}.

To address these issues, here we study the mixed valence regime of 
the generalized Anderson model at some particular values of 
the interactions where the model is exactly soluble (in the sense
that will be made precise below\cite{Bethe}). These points
in the interaction parameter space are the mixed valence counterparts
of the usual Toulouse point of the Kondo
problem\cite{Toulouse,AndersonYuval}, and are naturally
called the Toulouse points of the mixed valence problem.
We identify all the possible Toulouse points using the
bosonization approach. We construct an effective Hamiltonian,
and determine the single particle, spin-spin, 
and charge-charge correlation functions, at each of these
Toulouse points. There are two subtle aspects associated
with these Toulouse points. First, at the Toulouse points
some of the interactions are larger than the conduction
electron bandwidth. This makes the interpretation of these
Toulouse points subtle. We clarify this issue through a
comparison with an atomic expansion of the original impurity
Hamiltonian. Second, when applying the bosonization method
to the mixed valence problem, it turns out to
be essential to keep track of the anticommutation relation 
between the fermion fields. When properly understood, the
solutions to the generalized Anderson model at these
Toulouse points substantiate, and provide a simple physical
picture for, the mixed valent phases we have identified
through the renormalization group analysis.

Parallel to our previous works, an impurity problem defined by
the $U=\infty$ Anderson model with additional screening channels
have been studied using the numerical renormalization
group method\cite{Perakis}. Recently, two other groups\cite{Sire,Yulu}
have studied an exactly soluble point of that impurity
model. They have reached conclusions very different from ours,
and we will comment on the origin of these differences.

The setup of this paper is the following.  In Sec. \ref{sec:kondo}, we
review the construction of the Toulouse points in a context where the
physics is well understood, the Kondo model. In Sec. \ref{sec:mv}, we 
summarize the main results of our Coulomb gas analysis of the mixed
valence regime of the generalized Anderson model, and set up the 
formalism for our Toulouse-point analysis.
The Toulouse points are discussed in detail in the following sections.
The Toulouse points discussed in Sec. \ref{sec:sc1} and 
Sec. \ref{sec:sc2} lie deep in the strong coupling limit.
That of Sec. \ref{sec:intmed} describes the non-Fermi liquid,
intermediate phase. There is one more point, presented in Sec. 
\ref{sec:toulD}, which has an effective Hamiltonian 
corresponding to a rank-2 generalization of the 
Emery-Kivelson resonant level model\cite{EmeryKivelson}
associated with the two-channel Kondo problem. This effective 
Hamiltonian is not exactly soluble. We conclude with a 
comparison of our results with those of related works,
and a discussion about realizing the intermediate phase in
other models of strongly correlated electron systems. 
Two appendixes are included. Appendix \ref{sec:bos}
summarizes the bosonization procedure relevant to our
discussion, with an emphasis on the Klein factors
that keep track of the anticommutation relation
between fermions of different spins. Appendix \ref{sec:atomic}
substantiates the bosonization results presented in the main text
with those of an atomic analysis of the original Hamiltonian.

\bigskip
\section{\bf Toulouse points of the Kondo model}
\label{sec:kondo}

In this section, we introduce the notation, and review the 
bosonization method, in the context where the physics is 
well understood, the anisotropic one-channel spin$-{1 \over 2}$
Kondo model. 
We show that there are two values of $J_z$, the longitudinal component
of the Kondo exchange interaction, at which the problem is exactly 
soluble\cite{Bethe}. One of these is the well-known Toulouse 
point\cite{Toulouse}, for which the effective Hamiltonian is the 
resonant-level model\cite{Finkelstein}.
The other occurs at the infinite value for $J_{z}$. The effective
Hamiltonian for this second Toulouse point is the spin-boson 
Hamiltonian\cite{Blume,Leggett} close to a vanishing spin-boson
coupling.

The anisotropic Kondo problem describes a bath of spin$-{1 \over 2}$
conduction electrons coupled to an impurity which can
fluctuate between two states, $|\uparrow>$ and $|\downarrow>$.
The Hamiltonian is

\begin{eqnarray}
H = && \sum_{k \sigma} E_{k} c_{k\sigma}^{\dagger}c_{k\sigma}
+ H_{\perp} + H_{\parallel}\nonumber\\
H_{\perp} = &&\frac{J_{\perp}}{2} \sum_{\sigma} X_{\sigma \bar{\sigma}}
c^{\dagger}_{\bar{\sigma}} (0) c_{\sigma} (0) \nonumber\\
H_{\parallel} = &&\frac{J_{z}}{4} (\sum_{\sigma} \sigma X_{\sigma\sigma} )
(\sum_{\sigma^{\prime} }\sigma^{\prime}  c^{\dagger}_{\sigma^{\prime}}(0)
c_{\sigma^{\prime}} (0))
\label{hamkondo}
\end{eqnarray}
Here, $c_{k\sigma}^{\dagger}$ describes a free conduction electron 
bath with energy dispersion $E_k$, and 
$c_{\sigma}^{\dagger}(\vec{r}) = {1 \over \sqrt{N_{site}}}
\sum_{k} e^{i{\vec{k} \cdot \vec{r}}} c_{k\sigma}^{\dagger}$.
The impurity is located at $\vec{r}=0$, and is locally coupled to
the conduction electrons through an exchange interaction.
$J_{\perp}$ and $J_{z}$ are the transverse and longitudinal
components of this exchange interaction, respectively, and
are allowed to take different values. In Eq. (\ref{hamkondo}),
we have written the impurity spin operators in terms of
the Hubbard operators, $X_{\alpha\beta}=|\alpha><\beta|$, where
$\alpha,\beta$ describes the two impurity configurations,
$|\uparrow>$ and $|\downarrow>$. The following constraint

\begin{eqnarray}
X_{\uparrow\uparrow}+X_{\downarrow\downarrow}=1
\label{constraintkondo}
\end{eqnarray}
supplements Eq. (\ref{hamkondo}).

\subsection{Bosonization}
\label{sec:kondo-bos}

Given that the interaction occurs at $\vec{r}=0$ only, we need to
keep only the $S-$wave component of the conduction electrons.
This $S-$wave component is defined on the radial axis,
$r \in [0, +\infty)$, and can be further decomposed into an outgoing and
an incoming components. In a standard fashion,
we extend to the full axis, $x \in (-\infty, +\infty)$,
by retaining only one chiral component, which we denote by
$\psi_{\sigma} (x)$. We can then introduce a boson representation
for the $\psi_{\sigma} (x)$ field\cite{Emery}. The details of this 
procedure is given in Appendix \ref{sec:bos}. At the origin,

\begin{eqnarray}
\psi_{\sigma}^{\dagger} (0) = F_{\sigma}^{\dagger} {1\over \sqrt{2\pi a}}
{\rm e}^{i\Phi_{\sigma}}
\label{fermiono}
\end{eqnarray}
Here, $a$ is a cutoff scale which can be taken as a lattice spacing. 
$\Phi_{\sigma}$ is the shorthand notation for $\Phi_{\sigma}(x=0)$.
An important point is that, $\Phi_{\sigma}$ depends only on the 
$q\ne 0$ components of the Tomonaga bosons, $b_{q\sigma}$ and
$b_{q\sigma}^{\dagger}$. The operator $F_{\sigma}^{\dagger}$, and
its adjoint $F_{\sigma}$, are the so-called Klein factors.
They should be thought of as acting on the $q=0$ sector of the
Hilbert space for the Tomonaga bosons. More precisely, they can be defined
as the raising and lowering operators, respectively, in such a
Hilbert space\cite{Haldane81,Heid,Neuberg}. These operators
are unitary, and anticommute among the different spin species.
Furthermore, they commute with $b_{q\sigma}$ and $b_{q\sigma}^{\dagger}$
for $q \ne 0$ and, hence, also with $\Phi_{\sigma}$.

The Kondo Hamiltonian can be rewritten in the bosonized form,
$H=H_o+H_{\perp}+H_{\parallel}$, with,

\begin{eqnarray}
H_{o} = &&\frac{{v_F}}{4\pi} \int d x [(\frac{d\Phi_{s}}{dx})^{2} +
(\frac{d\Phi_{c}}{dx})^{2}]\nonumber\\
H_{\perp} = &&\frac{J_{\perp}}{4\pi a}
(X_{\uparrow\downarrow} F_{\downarrow}^{\dagger} F_{\uparrow}
e^{-i\Phi_{s} \sqrt{2}} + H.c.) \nonumber\\
H_{\parallel} = && \frac{\sqrt{2}\delta^s}{\pi\rho_o}
(X_{\uparrow \uparrow} - X_{\downarrow \downarrow})
(\frac{d\Phi_{s}}{dx})_{x=0} \frac{1}{2\pi}
\label{hamkondoboson}
\end{eqnarray}
where $\delta^s=\tan^{-1}(\pi \rho_o {J_z \over 4})$ is the phase 
shift associated with the potential ${J_z \over 4}$, $v_F$ the
Fermi velocity, and $\rho_o = {1 \over 2\pi v_F}$
the conduction electron density of states at the Fermi level.
In Eq. (\ref{hamkondoboson}), we have 
also introduced the charge and spin boson fields,
$\Phi_c (x) = {1 \over \sqrt{2}}(\Phi_{\uparrow} (x) + 
\Phi_{\downarrow} (x))$ and $\Phi_{s}(x) = {1 \over 
\sqrt{2}}(\Phi_{\uparrow} (x) - \Phi_{\downarrow} (x))$.

To construct soluble limits, we apply the following canonical 
transformation to the Hamiltonian,

\begin{eqnarray}
U = {\rm e} ^{-i \alpha \Phi_s (\sum_{\sigma}\sigma X_{\sigma\sigma})}
\label{canonicalt}
\end{eqnarray}
Using

\begin{eqnarray}
U^{+} \frac{d\Phi_{s}}{d x} U = &&\frac{d\Phi_{s}}{dx} - \alpha 2\pi
\delta(x) (\sum_{\sigma}\sigma X_{\sigma\sigma}) \nonumber\\
U^{+} X_{\uparrow\downarrow} U = &&e^{i2\alpha\Phi_{s}(o)}
X_{\uparrow\downarrow}
\label{canonicalt2}
\end{eqnarray}
the transformed Hamiltonian, $H_{eff}=U^+HU$, is

\begin{eqnarray}
H_{eff} &&=  H_o+  H_j+ \Delta H\nonumber\\
H_j && =  \frac{J_{\perp}}{4\pi a} [
X_{\uparrow\downarrow} F_{\downarrow}^{\dagger} F_{\uparrow}
e^{-i(\sqrt{2} -
  2\alpha)\Phi_{s}} + H.c.] \nonumber\\
\Delta H && = \frac{\tilde{\delta}^s}
{\pi\rho_o} 
(X_{\uparrow \uparrow} - X_{\downarrow \downarrow})
(\frac{d\Phi_{s}}{dx})_{x=0}\frac{1}{2\pi} 
\label{hamkondoct}
\end{eqnarray}
where $\tilde{\delta}^{s} \equiv \sqrt{2}\delta^{s}
- \pi \alpha$.
We choose an $\alpha$ such that what remains as the conduction electron
part in $H_j$ either has the form of a canonical 
fermion, or disappears entirely. The Toulouse points correspond to 
the bare values of the interactions such that
$\tilde{\delta}^{s} =0$, so that the residual interaction
$\Delta H$ vanishes.

\subsection{Toulouse Point I of the Kondo Problem}
\label{sec:kondo-toul}

The effective Hamiltonian assumes a simple form  when 
$\delta^s$ is close to $\delta^s_1= ({1 \over 2} - {\sqrt{2} \over 4})
\pi$, i.e., $J_z/4=J_z^1/4  ={1 \over {\pi \rho_o}} \tan^{-1} 
[({1 \over 2} - {\sqrt{2} \over 4})\pi]$. Choosing $\alpha = 
(\sqrt{2}-1)/2$ in Eq. (\ref{canonicalt}) leads to the following
transformed Hamiltnonian,

\begin{eqnarray}
H_{eff}^1= &&H_o + \frac{J_{\perp}}{2\sqrt{2\pi a}} (d^{\dagger} \eta
+H.c.)+\Delta H \nonumber\\
\Delta H  =&& \frac{\tilde{\delta}^{s}}{\pi\rho_o} 
(d^{\dagger}d-1/2)(\frac{d\Phi_{s}}{dx})_{x=0} {1 \over {2\pi}}
\label{reslevel}
\end{eqnarray}
where $\tilde{\delta}^{s}=\sqrt{2}(\delta^s-\delta^s_1)$ measures 
the deviation of the interaction $J_z$ from the chosen value $J_z^1$.
Here $\eta \equiv {1 \over \sqrt{2\pi a}}{\rm e}^{-i\Phi_s(0)} F_{\eta}$ 
is a canonical spin-less fermion field. $d^{\dagger} \equiv
X_{\uparrow\downarrow} F_{\downarrow}^{\dagger} F_{\uparrow} 
F_{\eta}^{\dagger}$, and its adjoint, $d\equiv
F_{\eta} F_{\uparrow}^{\dagger} F_{\downarrow}X_{\downarrow\uparrow}$,
satisfy the commutation relations appropriate for a fermion, 
$\{d,d^{\dagger}\}=X_{\uparrow\uparrow}+ X_{\downarrow\downarrow}=1$,
$\{d,d\}= \{d^{\dagger},d^{\dagger}\}=0$. 
The effective Hamiltonian, therefore, describes a resonant-level 
model\cite{Finkelstein}. The spin-spin correlation functions can
be calculated using the transformed longitudinal and transverse
spin operators,

\begin{eqnarray}
S^z_{eff}\equiv &&U^{+}({1 \over 2}\sum_{\sigma}\sigma X_{\sigma
\sigma})U={1 \over 2}\sum_{\sigma}\sigma X_{\sigma \sigma}
=d^{\dagger}d-{1 \over 2}\nonumber\\
S^+_{eff}\equiv &&U^{+}X_{\uparrow \downarrow}U =
e^{i(\sqrt{2}-1)\Phi_s} X_{\uparrow \downarrow} 
\label{Szxyct}
\end{eqnarray}
and noting that $X_{\uparrow \downarrow}$ has the same dimension as
${1 \over 2\pi a}F_{\uparrow}^{\dagger}F_{\downarrow}{\rm e}^{i\Phi_s}$.
The results are \cite{finiteT},

\begin{eqnarray}
<{\rm T}_{\tau} S^+(\tau) S^-(0)>&& = <{1 \over 2\pi a}
F_{\uparrow}^{\dagger} F_{\downarrow}
e^{i\sqrt{2}\Phi_s} (\tau) \nonumber\\
&&{1 \over 2\pi a}e^{-i\sqrt{2}\Phi_s} F_{\downarrow}^{\dagger}
F_{\uparrow}(0)>_{H_{eff}} \sim ({\rho_o \over \tau})^2 
\nonumber\\
<{\rm T}_{\tau} S^z(\tau) S^z (0)> && = <
(d^{\dagger}d-{1 \over 2})(\tau) \nonumber\\
&&(d^{\dagger}d-{1 \over 2})(0) >_{H_{eff}}
\sim ({\rho_o \over \tau})^2 
\label{spinspinkondotoul1}
\end{eqnarray}
This is the usual Toulouse point\cite{Toulouse,AndersonYuval,Hakim},
which describes the strong coupling, Fermi liquid fixed point.

\subsection{Toulouse Point II of the Kondo Problem}
\label{sec:kondo-toulII}

We now turn to the case of $\delta^s$ close to $\delta^s_2=
{1 \over 2}\pi$. This corresponds to an infinitely strong
antiferromagnetic interaction, $J_z^2=+\infty$. Choosing 
$\alpha = \sqrt{2}/2$ leads to the following terms for the 
transformed Hamiltonian, $H_{\perp} = \frac{J_{\perp}}{4\pi a} 
(X_{\uparrow\downarrow} F_{\downarrow}^{\dagger} F_{\uparrow} + H.c.)$,
and $\Delta H = \frac{\tilde{\delta}^{s}}{\pi\rho_o} 
(X_{\uparrow \uparrow} - X_{\downarrow \downarrow})
(\frac{d\Phi_{s}}{dx})_{x=0}{1 \over 2\pi}$, where
$\tilde{\delta}^s=\sqrt{2} (\delta^s-\delta^s_2)$. Introducing a 
pseudo-fermion operator, $f_{\sigma}^{\dagger}$,
and a pseudo-boson operator, $b_{\sigma}^{\dagger}$, through

\begin{eqnarray}
X_{\sigma\sigma^{\prime}} =&&
f_{\sigma}^{\dagger}f_{\sigma^{\prime}}\nonumber\\
b_{\sigma}^{\dagger} = &&f_{\sigma}^{\dagger}F_{\bar{\sigma}}^{\dagger}
\label{defbs}
\end{eqnarray}
we can rewrite the effective Hamiltonian as follows,

\begin{eqnarray}
H_{eff}^2= &&H_o - \frac{J_{\perp}}{4\pi a} (
b_{\uparrow}^{\dagger} b_{\downarrow} + H.c.)+\Delta H\nonumber\\
\Delta H  = && \frac{\tilde{\delta}^{s}}{\pi\rho_o} 
(b_{\uparrow}^{\dagger} b_{\uparrow} -
b_{\downarrow}^{\dagger} b_{\downarrow})
(\frac{d\Phi_{s}}{dx})_{x=0}{1 \over 2\pi}
\label{spinboson}
\end{eqnarray}
In terms of $b_{\sigma}^{\dagger}$, the constraint given in 
Eq. (\ref{constraintkondo}) can be rewritten as 
$\sum_{\sigma} b_{\sigma}^{\dagger} b_{\sigma} =1$. 
At $\delta^s=\delta^s_2$,
$H_{eff}^2$ describes a conduction electron bath decoupled from a spin
degree of freedom on which a magnetic field of strength $h \equiv
-\frac{J_{\perp}}{4\pi a}$ acts along the x direction in spin space.
In general, $H_{eff}^2$ describes the spin-boson Hamiltonian
with an Ohmic bath  introduced in Ref. \cite{Blume},
with $\frac{\tilde{\delta}^{s}}{\pi}$
being proportional to the square root of the dissipation parameter
$\alpha$ defined in the macroscopic quantum coherence
(MQC) context\cite{Leggett}.

Using the transformed spin operators,

\begin{eqnarray}
S^z_{eff}=&&{1 \over 2}(b_{\uparrow}^{\dagger} b_{\uparrow} -
b_{\downarrow}^{\dagger} b_{\downarrow})\nonumber\\
S^+_{eff}=&&-e^{i\sqrt{2}\Phi_s} F_{\uparrow}^{\dagger}F_{\downarrow}
b_{\uparrow}^{\dagger}b_{\downarrow}
\label{szxykondotoul2}
\end{eqnarray}
the transverse spin-spin correlation function is given by

\begin{eqnarray}
<{\rm T}_{\tau} S^+(\tau) S^-(o)>_{H} && \sim 
({\rho_o \over \tau})^2 
\label{kondotoultrans}
\end{eqnarray}
The calculation of the longitudinal spin-spin correlation function 
is somewhat more subtle. At $\tilde{\delta}^s=0$, there is only an 
oscillatory piece (in real time), with period $2h$. Expanding 
around this point, the leading non-vanishing non-oscillatory
term has the following long time behavior ($\tau \gg {1 \over | h |}$),

\begin{eqnarray}
<{\rm T}_{\tau} S^z(\tau) S^z(o)>_{H} \sim (\frac{\tilde{\delta}_s}
{\pi\rho_o h})^2 ({\rho_o \over \tau})^2
\label{kondotoullong}
\end{eqnarray}
In addition, the oscillatory piece damps out\cite{Leggett}, beyond 
a time scale of $\sim ({\pi \over \tilde{\delta}^{s} })^2 {1 \over | h |}$.
Therefore, the asymptotic long-time behavior of the longitudinal
spin-spin correlation function has the Fermi liquid,  
$({\rho_o \over \tau})^2$, form. This is the same form as that of the
transverse spin-spin correlation function. Such a long-time behavior
of the correlation functions is consistent with the physical picture
that, at long times the impurity spin degrees of freedom is ``merged''
with those of the conduction electron bath. 
The $({1 \over \tau})^2$ long-time behavior for the
longitudinal spin-spin correlation function in the spin-boson problem
is already known in the literature\cite{Spohn,Weiss,Chakravarty}.

Given that this last Toulouse point occurs at an infinitely strong
antiferromagnetic interaction, one might worry about the validity of
the bosonization approach. In Appendix \ref{sec:atomic}, we carry out
an atomic-expansion analysis for the Kondo Hamiltonian in the 
limit $J_{z} >> J_{\perp}, W$. The procedure is to first
diagonalize the $J_z$ coupling, taking $J_{\perp}=W=0$.
The lowest energy atomic configurations are 

\begin{eqnarray}
|1> = &&|\uparrow>_d |\downarrow>_0 \nonumber\\
|2> = &&|\downarrow>_d |\uparrow>_0
\label{atomickondo}
\end{eqnarray}
where the subscripts $d$ and $0$ label the impurity and the
Wannier orbital for the conduction electrons at the origin,
respectively. $J_{\perp}$ and $W$ couple these low energy configurations
with other higher energy ones. Integrating
out all the high energy configurations via  a canonical
transformation leads to the following effective Hamiltonian,

\begin{eqnarray}
H_{eff}= &&H_o^{\prime}+ J_{\perp} (X_{12} + X_{21}) \nonumber\\
&&+ J_z^{\prime} (X_{11} - X_{12}) \sum_{\sigma} 
\sigma c^{\dagger}_{1\sigma} c_{1\sigma}
\label{effkondoatomic}
\end{eqnarray}
where $c^{\dagger}_{1\sigma}$ creats a Wannier orbital of the
conduction electrons at the site nearest to the origin,
and $J_z^{\prime} \sim \frac{W^{2}}{J_z}$. 
The same canonical transformation also leads to the following
effective spin operators,

\begin{eqnarray}
(S^{z})_{eff} \sim && {1 \over 2} (X_{11} - X_{22}) \nonumber \\
(S^{+})_{eff} \sim &&  c^{\dagger}_{1\uparrow} c_{1\downarrow} X_{12}  
\label{szxykondoatomic}
\end{eqnarray}
These results from the atomic expansion, Eqs. (\ref{effkondoatomic})
and (\ref{szxykondoatomic}), are the direct analogs of
the bosonization results, Eqs. (\ref{spinboson}) and 
(\ref{szxykondotoul2}).

To summarize, we have emphasized two aspects associated with the
Toulouse points in the Kondo problem. First, we have explicitly
retained the Klein factors in the bosonization representation
of the fermion operators to keep track of the anticommutation
relations satisfied by fermions with different spins. For 
the Kondo problem {\it per se}, the single fermion operators
do not come into the Hamiltonian directly; only the fermion
bilinear operators do. Therefore, identical results could have been
derived without retaining these Klein factors. This is not true
for the mixed valence problem, where the single fermion operators
do appear  in the hybridization term. And, as will be shown in detail
in the following sections, it turns out to be essential to keep the
Klein factors explicitly when applying the bosonization method to
the mixed valence problem. Second, we discussed in detail a second 
Toulouse point that occurs at the infinite value for the
longitudinal component of the Kondo exchange interaction.
Through a comparison with the results of an expansion of the
atomic limit of the Kondo Hamiltonian, we established that,
even in this limit, bosonization can be applied. We also demonstrated
that the correct behaviors of the longitudinal spin-spin correlation
functions at this second Toulouse point can be derived by expanding
around the Toulouse point. These insights 
turn out to be quite useful to properly understand the Toulouse
points of the mixed valence problem.

\section{The Generalized Anderson Model}
\label{sec:mv}

In this section we define the generalized Anderson model, summarize 
our earlier scaling results, and set up the bosonization formalism 
for the Toulouse-point analysis appropriate for the generalized
Anderson model. The Hamiltonian of the generalized Anderson model is

\begin{eqnarray}
H= &&\sum_{k\sigma} E_k c_{k\sigma}^{\dagger}
c_{k\sigma} +\sum_{\sigma} E^o_d d_{\sigma}^{\dagger}
d_{\sigma}+{U \over 2} d^{\dagger}_{\uparrow} d_{\uparrow}
d^{\dagger}_{\downarrow} d_{\downarrow}\nonumber\\
&&+\sum_{\sigma} t ( d^{\dagger}_{\sigma} c_{\sigma} + h.c. )
+ {V } \sum_{\sigma,\sigma ' } d^{\dagger}_{\sigma} d_{\sigma}
c^{\dagger}_{\sigma ' } c_{\sigma '}
+V_p \sum_{\sigma} c^{\dagger}_{\sigma } c_{\sigma }
\nonumber\\
&&+ {J \over 4} \sum_{\sigma_1,\sigma_2,\sigma_3,\sigma_4}
{\bf \tau}_{\sigma_1\sigma_2} \cdot {\bf \tau}_{\sigma_3\sigma_4}
d^{\dagger}_{\sigma_1} d_{\sigma_2} c^{\dagger}_{\sigma_3 } c_{\sigma_4}
\label{hamil.us}
\end{eqnarray}
Here ${\bf \tau}$ label the Pauli matrices, and $\sigma = \uparrow ,
\downarrow $. To study the mixed valence regime, we focus on
$U=\infty$. The double occupancy configuration of the impurity
is then excluded. The three remaining configurations, $|0>$ and
$|\sigma> =d_{\sigma} ^{\dagger}|0>$, have
energies $E_0=0$ and $E_{\sigma}=E_d^o$, respectively. 
The hybridization $t$, the density-density interaction $V$, 
and the explicit spin exchange interaction $J$ describe 
the couplings between the impurity and the electron bath.
Anticipating the intrinsic particle-hole asymmetry in the
mixed valence regime, we have also allowed for a
potential scattering, $V_p$. This Hamiltonian is general
enough for the purpose of studying the interplay between
the local spin and charge fluctuations in the mixed
valence problem.

Following the procedure outlined in the previous section and given
in more detail in Appendix A, we can reduce the problem to that 
of an impurity coupled to a one-dimensional non-interacting 
conduction electron bath, with one chiral component
for each spin species, $\psi_{\sigma}(x)$. For the purpose of the
Coulomb gas as well as the Toulouse-point analyses, we also allow the
longitudinal and transverse components of the exchange interaction,
$J_z$ and $J_{\perp}$, to take different values. The
Hamiltonian can be rewritten as 

\begin{eqnarray}
H =&& H_o + E_d^o \sum_{\sigma}X_{\sigma \sigma} +H_{\perp t} 
+ H_{\perp j} +H_V\nonumber \\
H_{\perp t}=&&t \sum_{\sigma} ( X_{\sigma o} \psi_{\sigma}
+ H.c )\nonumber \\
H_{\perp j}=&&{ J_{\perp} \over 2}(X_{\uparrow \downarrow } 
\psi^{\dagger}_{\downarrow}(0)
\psi_{\uparrow}(0) + H.c ) \nonumber\\
H_V = &&\sum_{\alpha \sigma} V^{\sigma}_{\alpha} X_{\alpha\alpha}
\psi^{\dagger}_{\sigma} (0) \psi_{\sigma} (0)
\label{hamilmvboson}
\end{eqnarray}
Here, the impurity configuration $|\alpha>$ runs over $|0>$ and 
$|\sigma>$. This requirement amounts to the following constraint,
\begin{eqnarray}
X_{\uparrow \uparrow}
+X_{\downarrow \downarrow} +X_{oo} =1
\label{constraintmv}
\end{eqnarray}
$V_{\alpha}^{\sigma}$ denotes the local potential that the conduction 
electron of spin $\sigma$ experiences when the impurity is frozen
at the configuration $|\alpha>$. In terms of the parameters of 
Eq. (\ref{hamil.us}), 

\begin{eqnarray}
V_\sigma^\sigma &&=V + \frac{J_z}{4} + V_p \nonumber\\
V^{\bar{\sigma}}_{\sigma} &&= V -\frac{J_z}{4} +V_p \nonumber\\
V_{o}^{\sigma} &&=V_p
\label{potentials}
\end{eqnarray}

\subsection{The Coulomb Gas Representation and the Scaling Results }

The physics of this model can be understood by focusing on the
impurity degrees of freedom and tracing out the conduction
electrons. The partition function is then a sum over all
the histories of the impurity, each history being characterized
by a sequence of transitions of the impurity configurations
among $|0>$ and $|\sigma>$. This has been extensively 
discussed in our earlier papers\cite{SiKotliar1,SiKotliar2,KotliarSi}.
For completeness, we summarize the main results in this subsection.

The partition function has the following Coulomb gas form,

\begin{eqnarray}
{Z \over Z_0}
= \sum_{n=0}^{\infty} \sum_{[\alpha_1,...,\alpha_n;
\tau_1, ... \tau_n]}
{\rm exp}(-S[\alpha_1,...,\alpha_n;\tau_1, ... \tau_n])
\label{sumoverhis}
\end{eqnarray}
where

\begin{eqnarray}
S [\alpha_1,...,&&\alpha_n;\tau_1, ... \tau_n]
= \sum_{i<j} [K(\alpha_i, \alpha_j) + K(\alpha_{i+1}, 
\alpha_{j+1})\nonumber\\
&&- K(\alpha_i, \alpha_{j+1}) - K(\alpha_{i+1}, \alpha_{j}) ]
ln {(\tau_j - \tau_i) \over \xi}\nonumber\\
&&- \sum_i ln (y_{\alpha_i\alpha_{i+1}})
+\sum_{i}h_{\alpha_{i+1}} {(\tau_{i+1}-\tau_i) \over \xi}
\label{hisaction}
\end{eqnarray}
Here, $[\alpha_1,...,\alpha_n;\tau_1, ... \tau_n]$ labels a history,
with the impurity hopping quantum mechanically from one local state,
$|\alpha_i>$, to another, $|\alpha_{i+1}>$, at the (imaginary) time
$\tau_i$. The logarithmic interaction between the hopping events reflects
the reaction of the electron bath towards the changes of the
impurity configurations. Here, $\xi$ is the ultraviolet
inverse energy cutoff, and the strength of the logarithmic
interaction is characterized by the stiffness constants,
$\epsilon_t= -K(0,\sigma)$ and $\epsilon_j=-K(\sigma,\sigma'
\ne \sigma )$ which in turn are determined by the bare 
interaction strength of the original Hamiltonian.
Specifically,

\begin{eqnarray}
\epsilon_{t} = &&\frac{1}{2} [( 1 - \frac{\delta^{\sigma}_{\sigma} -
\delta^{\sigma}_{o}}{\pi})^2 + (\frac{\delta^{\bar{\sigma}}_{\sigma} -
\delta^{\bar{\sigma}}_{o}}{\pi} )^2] \nonumber\\
\epsilon_{j} = &&(1 - \frac{\delta^{\sigma}_{\sigma} -
\delta^{\bar{\sigma}}_{\sigma}}{\pi})^{2}
\label{epsilontj}
\end{eqnarray}
where 

\begin{eqnarray}
\delta_{\alpha}^{\sigma}=\tan^{-1}(\pi \rho_o V_{\alpha}^{\sigma})
\label{phaseshift}
\end{eqnarray}
is the phase shift that the conduction electron bath of spin $\sigma$ 
experiences when the impurity configuration is $|\alpha>$.
The fugacities $y_{\alpha,\beta}$ describe the
amplitudes for a quantum hopping between the configurations
$|\alpha>$ and $|\beta>$. More specifically, the charge fugacity
corresponds the hopping amplitude between two local states
with different charge quantum numbers and is proportional to the
hybridization amplitude, $y_{0,\sigma} \equiv y_t=t\xi$. Similarly,
the spin fugacity describes the hopping amplitude between two 
local states with different spin quantum numbers and 
is given by the transverse component of the exchange coupling,
$y_{\sigma, \sigma '} \equiv y_j = {J_{\perp} \over 2 } \xi$ for
$\sigma \ne \sigma^{\prime}$. Finally, the fields $h_{\alpha}$
describe the energy splittings among the local configurations.
In the absence of an external magnetic field, $h_0= - {2 \over 3}
E_d^o\xi$ and $h_{\sigma}= {1 \over 3} E_d^o\xi$. 

The physical content of this Coulomb gas representation is as follows.
$y_j$ and $\epsilon_j$ are the dimensionless quantities associated
with the transverse and longitudinal couplings of the usual spin
Kondo problem. $y_t$ and $\epsilon_t$, on the other hand, can be 
thought of as the analogous dimensionless quantities associated with
the transverse and longitudinal couplings in a charge Kondo problem.
And the difference between $h_{\sigma}$ and $h_o$, or rather the
impurity level, $E_d^o$, controls how ``soft'' the local charge 
fluctuations are; when $E_d^o$ is tuned through the conduction
electron Fermi level from far below to far above, the system
evolves from a local moment regime, through a mixed valence
regime, to the empty orbital regime. In the mixed valence regime,
Kondo-like processes in the spin channel and charge channel
are coupled together. The partition function given in Eq.
(\ref{hisaction}) is the proper generalization of Haldane's
Coulomb gas representation of the mixed valence problem
such that the interplay between the spin and charge channels
are incorporated systematically. 

Our renormalization group analysis of this partition function
establishes the existence of three, and only three, mixed valence
phases:  1) the usual strong coupling phase. Both the
spin and charge Kondo problems are in the strong coupling regime;
rapid fluctuations between all three local configurations take
place and the conduction electrons quench both the charge and
spin degrees of  freedom of the impurity; 2) a weak coupling
phase where neither the local charge nor the local spin degrees
of freedom is quenched. Both the spin and charge Kondo 
problems are in the weak coupling regime, and all three
atomic configurations decouple asymptotically at low energies;
3) an intermediate phase where the local spin degrees of freedom
is quenched, but the local charge degrees of freedom is not.
Here, the charge Kondo problem is in the weak coupling regime
despite of the fact that the spin Kondo problem is 
in the strong coupling regime.  There are {\it two} local 
configurations carrying different charges which are
decoupled asymptotically. The phase diagram is properly
given in terms of the $\epsilon_t-\epsilon_j$ parameter space.
The strong coupling phase occurs when $\epsilon_t < 1$, as well as
over a range of $\epsilon_j < 1$ when $\epsilon_t >1$.
The weak coupling phase can occur only when both
$\epsilon_t > 1$ and $\epsilon_j > 1$. From Eq.
(\ref{epsilontj}) this condition means that the effective
exchange interaction has to be ferromagnetic.
The weak coupling phase is, therefore, very likely to be
of only academic value. The intermediate phase arises
over a range within $\epsilon_t > 1$ and $\epsilon_j < 1$.
For the model Hamiltonian (\ref{hamil.us}) this means
an antiferromagnetic exchange interaction, and a finite
attractive density-density interaction, $V$, between
the impurity and conduction electrons. As will be further
discussed in Sec. \ref{sec:realistic}, taking $V$ as 
an effective parameter this condition can be satisfied
in a variety of realistic models. 
The transition between the different regimes is analogous
to the localization phase transition studied in the
context of MQC\cite{Leggett} and more recently in
the context of transport through constrictions in interacting
quantum wires\cite{KaneFisher}.

\subsection{Construction of the Toulouse
Points of the Mixed Valence Problem}

We now proceed to derive the Toulouse points. Bosonizing the 
conduction electrons, and making a canonical transformation using

\begin{eqnarray}
U = e^{-i\alpha\Phi_{s} \sum_{\sigma}\sigma X_{\sigma\sigma}}
e^{-i \beta \Phi_{c} (\sum_{\sigma}X_{\sigma\sigma}-X_{oo})}
e^{-i \gamma \Phi_{c}}
\label{canontransmv}
\end{eqnarray}
we arrive at the effective Hamiltonian, $H_{eff}=U^{+}HU$,

\begin{eqnarray}
H_{eff} &&=  H_o+  E_d^o \sum_{\sigma}X_{\sigma\sigma} 
+H_t + H_j+ \Delta H\nonumber\\
H_t &&= \frac{t}{\sqrt{2\pi a}} \sum_{\sigma} [X_{\sigma
  o} F_{\sigma} e^{-i(\frac{1}{\sqrt{2}} - 2\beta)\Phi_{c} }
e^{-i\sigma(\frac{1}{\sqrt{2}} - \alpha)\Phi_{s}} +  H.c.] \nonumber \\
H_j && =  \frac{J_{\perp}}{4 \pi a} [X_{\uparrow\downarrow} 
F_{\downarrow}^{\dagger}F_{\uparrow} e^{-i(\sqrt{2} -
  2\alpha)\Phi_{s}} + H.c.] \nonumber\\
\Delta H && = \sum_{\alpha} X_{\alpha\alpha} [
\frac{\tilde{\delta}_{\alpha}^s}{\pi\rho_o} (\frac{d\Phi_{s}}{dx})_{x=0} 
{1 \over 2\pi}
+ {\tilde{\delta}_{\alpha}^c \over \pi \rho_o}
(\frac{d\Phi_{c}}{dx})_{x=0} {1 \over 2\pi} ]
\label{canontransfmv}
\end{eqnarray}
The notations are the same as in the previous section.
$\tilde{\delta}^{s}_{\sigma} \equiv \delta^{s}_{\sigma} - {\pi}
\alpha\sigma$, $\tilde{\delta}^{c}_{o} \equiv \delta^{c}_{o} + 
\pi(\beta-\gamma)$, and $\tilde{\delta}^{c}_{\sigma} \equiv 
\delta^{c}_{\sigma} - \pi (\beta+\gamma)$. Here, 
$\delta^{c}_{\alpha} \equiv \frac{1}{\sqrt{2}} \sum_{\sigma}
\delta^{\sigma}_{\alpha}$ and $\delta^{s}_{\alpha} \equiv
\frac{1}{\sqrt{2}} \sum_{\sigma} \sigma \delta^{\sigma}_{\alpha}$,
where $\delta^{\sigma}_{\alpha}$ is defined in Eq. (\ref{phaseshift});
more specifically,

\begin{eqnarray}
\delta^{s}_{\sigma} \equiv &&\sigma (\delta_{\sigma}^{\sigma}
- \delta_{\sigma}^{\bar{\sigma}})/\sqrt{2}\nonumber\\
\delta^{c}_{\sigma} \equiv &&(\delta_{\sigma}^{\sigma}
+ \delta_{\sigma}^{\bar{\sigma}})/\sqrt{2}\nonumber\\
\delta^{c}_{o} \equiv &&(\delta_{o}^{\sigma}
+ \delta_{o}^{\bar{\sigma}})/\sqrt{2} =\sqrt{2}\delta_o^{\sigma}
\label{deltasc}
\end{eqnarray}
The last equality of the last equation holds in the absence
of an external magnetic field. The latter also
ensures that $\delta_o^{s}=0$. 

To construct a Toulouse point, we choose the parameters $\alpha$
and $\beta$ such that the the conduction electron operators appearing
in the transformed $H_t$ and $H_j$ have simple forms.
The Toulouse point then corresponds to choosing the
bare values of the interactions such that the transformed $\Delta H$
vanishes. This means taking 
$\delta^{s}_{\sigma} = \sigma \pi \alpha $, $\delta^{c}_{\sigma}
= \pi (\beta+\gamma)$, and $\delta^{c}_{o} = \pi(\gamma-\beta)$. 
From Eq. (\ref{deltasc}), these conditions are equivalent to
requiring that the phase shifts $\delta_{\alpha}^{\sigma}$
take the following values,

\begin{eqnarray}
\delta^{\sigma}_{\sigma} && = {1 \over \sqrt{2}} (
\beta + \gamma + \alpha ) \pi \nonumber\\
\delta^{\bar{\sigma}}_{\sigma} &&= {1 \over \sqrt{2}} (
\beta + \gamma - \alpha ) \pi \nonumber\\
\delta^{\sigma}_{o} &&= {1 \over \sqrt{2}} (
\gamma - \beta ) \pi 
\label{phaseshifttoul}
\end{eqnarray}
According to Eq. (\ref{epsilontj}), the Coulomb gas stiffnesses 
at the Toulouse point are

\begin{eqnarray}
\epsilon_{t} = &&\frac{1}{4} [( \alpha + 2 \beta -\sqrt{2})^2+
(\alpha-2\beta)^2]\nonumber\\
\epsilon_{j} = &&(\sqrt{2}\alpha -1 )^2
\label{epsilontjtoul}
\end{eqnarray}

Eq. (\ref{phaseshifttoul}), together with Eqs. (\ref{potentials}) and
(\ref{phaseshift}), allow us to determine the parameters of the
generalized Anderson model at the Toulouse points. In terms of 
the phase shifts, they are

\begin{eqnarray}
\delta(J_z/4) \equiv &&\tan^{-1}[{1\over 2}\tan(\delta_{\sigma}
^{\sigma})-{1\over 2}\tan (\delta_{\sigma}
^{\bar{\sigma}})] \nonumber\\
=&&\tan^{-1}[{1 \over 2}\tan(\pi{ {\beta + \gamma + \alpha}
\over \sqrt{2}})- {1 \over 2}\tan (\pi{{\beta + \gamma - \alpha}
\over \sqrt{2}})]\nonumber\\
\delta(V) \equiv && \tan^{-1}[{1 \over 2}\tan(\delta_{\sigma}
^{\sigma})+{1\over 2}\tan (\delta_{\sigma}^{\bar{\sigma}})
-\tan (\delta_o^{\sigma})]\nonumber\\
=&&\tan^{-1}[{1\over 2}\tan(\pi{{\beta + \gamma + \alpha}
\over \sqrt{2}})\nonumber\\
&&+{1\over 2}\tan (\pi{{\beta + \gamma - \alpha} \over
\sqrt{2}})-{1 \over 2} \tan ( \pi { {\gamma -\beta}
\over \sqrt{2}})]\nonumber\\
\delta(V)  = &&\pi {{\gamma -\beta}\over \sqrt{2}}
\label{parameterstoul}
\end{eqnarray}

At a Toulouse point, the correlation functions can be calculated in 
terms of the single particle, off-diagonal spin, diagonal spin, and
density operators in the transformed basis,

\begin{eqnarray}
(d_{\sigma}^{\dagger})_{eff} \equiv &&U^{\dagger}X_{\sigma o}U = 
e^{i\alpha \Phi_s \sigma} e^{2i\beta \Phi_c} X_{\sigma o} \nonumber\\
S^+_{eff}\equiv &&U^{\dagger}X_{\uparrow \downarrow}U =
e^{2i\alpha\Phi_s} X_{\uparrow \downarrow} \nonumber\\
S^z_{eff}\equiv &&U^{\dagger}({1 \over 2}\sum_{\sigma}\sigma X_{\sigma
\sigma})U={1 \over 2}\sum_{\sigma}\sigma X_{\sigma \sigma}\nonumber\\
\rho_{eff}\equiv &&U^{\dagger} (\sum_{\sigma}X_{\sigma \sigma})U=
\sum_{\sigma}X_{\sigma \sigma}
\label{transopmv}
\end{eqnarray}

We found that only three independent points exist where the transformed
Hamiltonian is exactly soluble, and one more point where the effective 
Hamiltonian assumes a simple form but is not exactly soluble. Before
proceeding to analyze each of these points, we end this section with
a comment on the role of the $\gamma$ term in the canonical
transformation and, likewise, the role of the potential scattering
term in the generalized Anderson model. In general, Eq. 
(\ref{phaseshifttoul}) may require that, at the Toulouse points, one or
several of the phase shifts, $\delta_{\alpha}^{\sigma}$, be outside 
the range specified by the unitarity limit, $[-{\pi \over 2},
{\pi \over 2}]$. While bound states can lead to the violation of 
the unitarity bounds of the coupling constants\cite{Hakim},
in our case we can take advantage of the freedom of varying the strength
of the potential scattering term to make all the phase shifts
in Eq. (\ref{phaseshifttoul}) to fall in the range $[-{\pi \over 2},
{\pi \over 2}]$. This will be seen to be especially convenient for the
purpose of demonstrating the consistency of the bosonization results
with those of the atomic expansions shown in Appendix \ref{sec:atomic}.
The physics is independent of the potential scattering term,
as can be clearly seen from
the $\gamma-$independence of the effective Hamiltonian,
Eq. (\ref{canontransfmv}), as well as the $\gamma-$independence
of the Coulomb gas stiffnesses at the Toulouse points,
Eq. (\ref{epsilontjtoul}).

\bigskip
\section{\bf Strong Coupling Toulouse Point I}
\label{sec:sc1}

The first Toulouse point arises with the choice of 
$\alpha = \frac{\sqrt{2}}{2}$ and $\beta = \frac{\sqrt{2}}{4}$.
At this point, Eq. (\ref{epsilontjtoul}) gives $\epsilon_t=0$
and $\epsilon_j=0$. The Coulomb gas analysis implies that
we are deep in the strong coupling regime. The requirement
that all the phase shifts $\delta_{\alpha}^{\sigma}$ are
within the range $[-{\pi \over 2},{\pi \over 2}]$ leads to
a unique choice for $\gamma=-{\sqrt{2} \over 4}$. With this
choice of $\gamma$, $\delta_{\sigma}^{\sigma}=\pi/2$,
$\delta_{\sigma}^{\bar{\sigma}} = -\pi/2$,
and $\delta_o^{\sigma} =-\pi/2$. Equivalently, $ \delta (J_{z}/4)$
and $ \delta (V)$ are both equal to $\pi/2$, while
$\delta(V_p)=-\pi/2$. This corresponds to an infinite
antiferromagnetic exchange interaction and an infinite
repulsive density-density interaction. 

In this case, $H_t= \frac{t}{\sqrt{2\pi a}} \sum_{\sigma}
(X_{\sigma o}F_{\sigma}+ H.c.)$, and $H_j = 
\frac{J_{\perp}}{4\pi a} (X_{\uparrow\downarrow} F_{\downarrow}
^{\dagger} F_{\uparrow}+ H.c.)$. We now introduce pseudofermion
creation and annihilation operators $f_{\sigma}^{\dagger}$ and
$f_{\sigma}$, and pseudoboson creation and annihilation operators
$b_{\sigma}^{\dagger}$ and $b_{\sigma}$, as in Eq. (\ref{defbs}).
We further introduce the pseudoboson creation and annihilation 
operators $b_{o}^{\dagger}$ and $b_{o}$ such that
$X_{\sigma o} = f_{\sigma}^{\dagger}b_o$ and
$X_{o o} = b_{o}^{\dagger}b_o$. The effective Hamiltonian
becomes

\begin{eqnarray}
H_{eff}^A = &&H_o + E_d^o\sum_{\sigma} b_{\sigma}^{\dagger}b_{\sigma}
+\frac{t}{\sqrt{2\pi a}}\sum_{\sigma} (b_{\sigma}^{\dagger} b_o+
b_o^{\dagger} b_{\sigma}) \nonumber\\
&&- \frac{J_{\perp}}{4\pi a} 
(b_{\uparrow}^{\dagger} b_{\downarrow} + H.c.)+\Delta H 
\nonumber\\
\Delta H = &&({\kappa_c \over 2\pi\rho_o}) 
(\sum_{\sigma} b_{\sigma}^{\dagger}b_{\sigma} -
b_{o}^{\dagger}b_{o}) ({d \Phi_c \over dx})_{x=0}
{1 \over 2\pi}\nonumber\\
&&+({\kappa_s \over 2\pi\rho_o}) 
(\sum_{\sigma} \sigma b_{\sigma}^{\dagger}b_{\sigma} )
({d \Phi_s \over dx})_{x=0}
{1 \over 2\pi}
\label{effha}
\end{eqnarray}
And the constraint Eq. (\ref{constraintmv}) is rewritten as

\begin{eqnarray}
\sum_{\sigma} b_{\sigma}^{\dagger}b_{\sigma} 
+ b_{o}^{\dagger}b_{o}  = 1
\label{constraintmv1}
\end{eqnarray}
In deriving the effective Hamiltonian, we have chosen 
$\delta_0^{\sigma}$ such that the transformed potential 
scattering term vanishes. With this choice, $\kappa_c$ and 
$\kappa_s$ can be explicitly written in terms of the original
phase shifts as $\kappa_c=\sqrt{2}(\delta_{\sigma}^{\sigma} 
+\delta_{\sigma}^{\bar{\sigma}})$ and $\kappa_s=\sqrt{2}
(\delta_{\sigma}^{\sigma}-\delta_{\sigma}
^{\bar{\sigma}}-\pi)$. They vanish at the Toulouse point.

At the Toulouse point, the Hamiltonian describes a three-level
system decoupled from the conduction electrons. It is the 
three-level analog of the two level spin-boson 
Hamiltonian with zero dissipation. The three-level system
can be diagonalized exactly, with three eigenstates,
$|s0> = u |A> + v |o>$, $|t>=|B>$, and $|s1> = -v |A> + u |o>$.
Here,

\begin{eqnarray}
|A>\equiv && {1 \over \sqrt{2}} (b_{\uparrow}^{\dagger}
+b_{\downarrow}^{\dagger})|vac> \nonumber\\
|B>\equiv && {1 \over \sqrt{2}} (b_{\uparrow}^{\dagger}
-b_{\downarrow}^{\dagger})|vac> \nonumber\\
|o>\equiv && b_{o}^{\dagger} |vac >
\label{newbasismv1}
\end{eqnarray}
define a new basis for the atomic states.
In the absence of hybridization, these three configurations have
energies $E_A= E_d^o-\frac{J_{\perp}}{4\pi a}$, 
$E_B = E_d^o + \frac{J_{\perp}}{4\pi a}$, 
and $E_o=0$. The hybridization term mixes $|A>$ and $|o>$,
leading to the bonding and anti-bonding states
$|s0>$ and $|s1>$. $u$ and $v$ are the coherence factors,
$u^2=1-v^2 ={1 \over 2} (1 -{ {E_d^o-J_{\perp} /{4\pi a}}
\over \sqrt{(E_d^o-J_{\perp}/{4\pi a}})^2 + 8(t/\sqrt{2\pi a})^2})$. 
The energies of the eigenstates are
$E_{s0} = [(E_d^o - J_{\perp}/{4\pi a})^2 + 8(t/\sqrt{2\pi a})^2]/2$, 
$E_{t} = E_d^o + J_{\perp}/{4\pi a}$, and $E_{s1}
= [(E_d^o - J_{\perp}/{4\pi a})
+ \sqrt{(E_d^o -J_{\perp}/{4\pi a})^2 + 8(t/\sqrt{2\pi a})^2}]/2$.
Given that $J_{z}$ is antiferromagnetic, $J_{\perp}$ should also be
antiferromagnetic (positive). Therefore, irrespective of
$E_d^o$, $|s0>$ is always the ground state. 

The ground state of the whole system is $|gs>=U^{+}|s0>|FS>$.
The correlation functions can be calculated from the transformed
single particle, spin and density operators 

\begin{eqnarray}
(d_{\sigma}^{\dagger})_{eff} = &&e^{i\sigma \Phi_s/\sqrt{2}}
e^{i\Phi_c/\sqrt{2}} {1 \over \sqrt{2}}(X_{Ao}
+\sigma X_{Bo})F_{\sigma}^{\dagger}\nonumber\\
S^+_{eff} = &&e^{i\sqrt{2}\Phi_s} {1 \over 2}(X_{AA}-X_{BB}
+X_{BA}-X_{AB})\nonumber\\
S^z_{eff} = &&{1 \over 2}(X_{AB}+X_{BA})\nonumber\\
\rho_{eff}= &&(X_{AA}+X_{BB}-X_{oo})
\label{transopmv1}
\end{eqnarray}
The single-particle and the transverse spin susceptibility are
straightforward to calculate and have the Fermi liquid forms,

\begin{eqnarray}
<{\rm T}_{\tau}d_{\sigma} (\tau ) d_{\sigma}^{\dagger} (0)>
&& \sim {\rho_o \over  \tau }\nonumber\\
<{\rm T}_{\tau}S^-(\tau) S^+ (0) > && \sim (\frac{\rho_o}{\tau})^2
\end{eqnarray}
At the Toulouse point, the longitudinal spin-spin correlation
function and the density-density correlation function again
have the oscillatory behaviors in real time at the
Toulouse point. As for the longitudinal spin correlation
function near the Toulouse point II of the Kondo problem,
a direct expansion in terms of the deviation from the
Toulouse point leads to the following Fermi liquid
behaviors for these two correlation functions,

\begin{eqnarray}
<{\rm T}_{\tau} S^z(\tau) S^z (0) > && \sim ({\kappa_s 
\over 2\pi\rho_oh_s})^2 (\frac{\rho_o}{\tau})^2\nonumber\\
<{\rm T}_{\tau}\rho(\tau) \rho (0) > && \sim ({\kappa_c 
\over 2\pi\rho_oh_c})^2 (\frac{\rho_o}{\tau})^{2}
\end{eqnarray}
where $h_s=J_{\perp} /4\pi a$ and $h_c = t /\sqrt{2\pi a}$.
This Toulouse point, therefore, describes a strong coupling,
Fermi liquid state.

Unlike for the Kondo problem, keeping track of the anticommutation
relation between fermions of different spins in the boson
representation plays an essential role. Failing to do that,
the sign of $H_j$ term would be reversed,
leading to the atomic configuration $|A>$ having energy
$E_d^o+{J_{\perp} \over 4\pi a}$ instead of
$E_d^o-{J_{\perp} \over 4\pi a}$, and $|B>$ having
energy $E_d^o-{J_{\perp} \over 4\pi a}$ instead of
$E_d^o + {J_{\perp} \over 4\pi a}$. In that case,
the spin-flip exchange interaction ($J_{\perp}$) would
make the configuration $|B>$ energetically more favorable
than $|A>$, while the hybridization ($t$) term, which mixes
$|A>$ with $|o>$, would favor $|s0>$ instead of $|t>=|B>$.
This competition between the spin-exchange and hybridization would
then lead to a level-crossing as $E_d^o$ is varied: the ground state
changes from $|t>$ to $|s0>$ as the $d-$level varies from
far below to far above the Fermi level. 
The level-crossing found in earlier works on related
problems\cite{Sire,Yulu} results
from this failing to keep track of the anticommutation
relation between the bosonized fermion fields.

We can gain much physical insights into our results by determining
the quantum numbers of the respective impurity eigenstates. 
Within the bosonization approach, the meaning of the atomic
configurations, $|A>$, $|B>$ and $|o>$, is somewhat obscure.
The physical content of these configurations becomes 
transparent once we compare them with the atomic configurations
that appear in a perturbation expansion of the original Hamiltonian
in terms of $J_{\perp}/J_z$, $J_{\perp}/V$,  $t/J_z$, $t/V$, $W/J_z$,
and $W/V$. This atomic expansion is a natural procedure given that
$J_z, V \gg  J_{\perp},t,W$ at the Toulouse point. The details
of this atomic expansion can be found in Appendix \ref{sec:atomic}.
From Eqs. (\ref{atomicmv1}) and (\ref{effHmv1}), 
$|A>$, $|B>$ and $|o>$ are identified with

\begin{eqnarray}
|A> &&= {1 \over \sqrt{2}}(|\uparrow>_d |\downarrow>_0
- |\downarrow>_d |\uparrow>_0)\nonumber\\
|B> &&= {1 \over \sqrt{2}}(|\uparrow>_d |\downarrow>_0
+ |\downarrow>_d |\uparrow>_0)\nonumber\\
|O> &&= |0>_d|2>_0
\label{atomic-config1}
\end{eqnarray}
$|A>$ is the local singlet formed between the impurity spin
and the conduction electron spin at the impurity site, 
$|B>$ the $S_z=0$ state of the local triplet,
and $|o>$ the singlet with the impurity empty of
electrons and the local conduction electron orbital
doubly occupied. It is clear that both the exchange
interaction ($J_{\perp}$) term and the hybridization
($t$) term favor the same singlet state $|s0>=u|A>+v|o>$,
for arbitrary values of $E_d^o$. No level crossing occurs.

\bigskip
\section{\bf Strong Coupling Toulouse Point II}
\label{sec:sc2}

We now choose $\alpha = \frac{\sqrt{2}}{2}$ and $\beta =
{{\sqrt{2}-2 } \over 4}$. The corresponding Coulomb gas
stiffnesses are $\epsilon_t=1/2$ and $\epsilon_j=0$, and
the Coulomb gas analysis would again predict that the
system is in the strong coupling phase. As in the
previous section, the requirement that all the phase
shifts fall in the range $[-{\pi \over 2}, {\pi \over 2}]$
leads to a unique choice for $\gamma={1 \over 2}
-{\sqrt{2} \over 4}$. This choice corresponds to $\delta_{\sigma}
^{\sigma} =\pi/2$, $\delta_{\sigma} ^{\bar{\sigma}} = -\pi/2$, 
and $\delta_o^{\sigma}={{\sqrt{2}-1} \over 2}\pi$.
Equivalently, $ \delta (J_{z}/4)=\pi/2$, $ \delta ( V)=-
{{\sqrt{2}-1}\over 2}\pi$, and $\delta(V_p)={{\sqrt{2}
-1}\over 2}\pi$. We have an infinite antiferromagnetic
exchange interaction and a large, but finite, attractive
density-density interaction.

In this case, $H_t=\frac{t}{\sqrt{2\pi a}} \sum_{\sigma}
(X_{\sigma o} F_{\sigma} e^{-i\Phi_{c}} + H.c.)$,
and $H_j=\frac{J_{\perp}}{4\pi a} (X_{\uparrow\downarrow} 
F_{\downarrow}^{\dagger}F_{\uparrow}+H.c.)$. Introducing
a fermion field, $\eta^{\dagger} =F_{\eta}^{\dagger}
{1 \over \sqrt{2\pi a}}{e^{i\Phi_{c}}}$, and defining
$f_{\sigma}^{\dagger} = X_{\sigma 0} F_{\sigma} F_{\eta}
^{\dagger}$, we derive the following effective
Hamiltonian,

\begin{eqnarray}
H_{eff}^B = &&\sum_{k} E_{k}\eta^{+}_{k}\eta_{k} + 
 E_{d}^o f_{\sigma}^{\dagger}f_{\sigma}+
t[(\sum_{\sigma}f_{\sigma}^{\dagger}) \eta + H.c ]\nonumber\\
&&- \frac{J_{\perp}}{4\pi a}(f_{\uparrow}^{\dagger}f_{\downarrow}
+H.c.)
\label{effhamB}
\end{eqnarray}
It is convenient to introduce a new basis for the
atomic configurations, $|A>={1 \over \sqrt{2}}(f_{\uparrow}
^{\dagger} + f_{\downarrow}^{\dagger})|0>$, 
$|B>={1 \over \sqrt{2}} (f_{\uparrow}^{\dagger} -f_{\downarrow}
^{\dagger})|0>$, and $|o>$. The corresponding atomic levels
are $E_A= E_{d}^o -\frac{J_{\perp}}{4\pi a}$,
$E_B= E_{d}^o +\frac{J_{\perp}}{4\pi a}$,
and $E_o = 0$. In this new basis, the constraint becomes 

\begin{eqnarray}
X_{AA} + X_{BB} + X_{oo} = 1
\label{constraintmvB}
\end{eqnarray}
and the effective Hamiltonian can be rewritten as

\begin{eqnarray}
H_{eff}^B= &&\sum_{k} E_{k} \eta^{\dagger}_{k} \eta_{k} + \sqrt{2} t
( X_{Ao} \eta + H.c.)\nonumber\\
&&+[E_{d}^o -\frac{J_{\perp}}{4\pi a}] X_{AA} 
+ [E_{d}^o + \frac{J_{\perp}}{4\pi a}] X_{BB}
\label{effhamBrl}
\end{eqnarray}

Among the three impurity configurations $|A>$, $|B>$, and $|o>$,
$|A>$ and $|o>$ hybridize with the conduction electrons, while
$|B>$ is decoupled. Therefore, this Hamiltonian can be
diagonalized exactly. The Hilbert space factorizes into two
sectors. Again, since $J_z$ is antiferromagnetic, $J_{\perp}$
should also be taken as antiferromagnetic. It is clear that,
the low lying excitations lie in the sector in which the ground
state is

\begin{eqnarray}
\phi = |o> \phi_o + |A> \phi_{A}
\end{eqnarray}
where $\phi_{o}$ and $\phi_{A}$ are the conduction electron
wave functions such that $\phi$ is the ground state of
a resonant level model with $X_{Ao}$ treated as a free
fermion operator. The ground state energy (relative to
a free Fermi sea) is that of the resonant-level model
with a resonance width $2\pi\rho_o t^{2}$ and an
effective $d-$level $E_{d}^o - \frac{J_{\perp}}{4\pi a}$.

Using $d_{\sigma}^{\dagger} = e^{i\sqrt{2}\Phi_s\sigma/2} 
e^{i(\sqrt{2}-2)\phi_c/2} X_{\sigma o}$, 
$S^{+} = e^{i\sqrt{2}\Phi_s} X_{\uparrow \downarrow}$,
$S^{z} = (X_{AB} + X_{BA})/2$, and 
$\rho = (X_{AA}+X_{BB}- X_{oo})$, and noticing that 
$X_{Ao}$ has the dimension of ${1 \over {\sqrt{2\pi a}}} e^{i\phi_c} $,
we find that all the correlation functions have the Fermi liquid form,

\begin{eqnarray}
<{\rm T}_{\tau}d_{\sigma} (\tau) d_{\sigma}^{\dagger} (0)> && \sim 
{\rho_o \over \tau}\nonumber\\
<{\rm T}_{\tau}S^{-} (\tau) S^{+} (0)> && \sim (\frac{\rho_o}
{\tau})^2\nonumber\\
<{\rm T}_{\tau} S^z (\tau) S^z (0)> && \sim ({ \kappa_s 
\over 2\pi \rho_oh_s } )^2 (\frac{\rho_o}{\tau})^2\nonumber\\
<{\rm T}_{\tau} \rho (\tau) \rho (0)> && \sim (\frac{\rho_o}{\tau})^2
\label{transopmv2}
\end{eqnarray}
where $\kappa_s$ is again the deviation from the Toulouse point. 
This establishes the strong coupling, Fermi liquid nature of
this Toulouse point.

Except for the change of sign in $J_{\perp}$, the effective Hamiltonian
(\ref{effhamB}) is identical to that of Refs. \cite{Sire} and 
\cite{Yulu}.
The sign change for $J_{\perp}$, however, plays a crucial role.
It implies the absence of a level-crossing as a function of 
$E_d^o$. Similar to what happens in the previous subsection,
the result is the Fermi liquid behavior for the longitudinal
spin-spin correlation function, as for the transverse spin-spin
correlation function. No unexpected fixed points as discussed in
Refs. \cite{Sire} and \cite{Yulu} occur. The bosonization 
results are consistent with our previous renormalization group results.

\bigskip 
\section{\bf A Toulouse Point for the Intermediate Phase}
\label{sec:intmed}

In Refs. \cite{SiKotliar1,SiKotliar2,KotliarSi}
we found a new mixed valence phase, which we called the
intermediate phase. It is a non-Fermi liquid with quasiparticle-like
spin excitations and incoherent charge excitations.
In this section, we present 
a Toulouse point which clearly exhibits the physics of the
intermediate phase. It occurs with the choice of $\alpha =
\frac{\sqrt{2}}{2}$ and $\beta = -\frac{\sqrt{2}}{4}$.
The corresponding Coulomb gas stiffnesses are $\epsilon_t=1$
and $\epsilon_j=0$. While it is not possible to determine
the precise boundary between the intermediate phase and
the strong coupling phase, it is not inconsistent with the
Coulomb gas results that these values of the Coulomb gas
stiffnesses lie close to such a boundary. Taking
$\gamma={\sqrt{2} \over 4}$ specifies the phase shifts at the
Toulouse point,  $\delta_{\sigma}^{\sigma}={\pi \over 2}$,
$\delta_{\sigma} ^{\bar{\sigma}} =-{\pi \over 2}$ and 
$\delta_o^{\sigma} = {\pi \over 2}$. Equivalently, 
$\delta(J_z) = \frac{\pi}{2}$, $\delta (V)=- \frac{\pi}{2}$, and 
$\delta (V_p)= \frac{\pi}{2}$. These parameters correspond to an
infinite antiferromagnetic exchange interaction and an infinite
attractive density-density interaction.

In the canonically-transformed Hamiltonian, $H_t = \frac{t}
{\sqrt{2\pi a}}\sum_{\sigma} [X_{\sigma o} F_{\sigma} 
e^{-i\Phi_{c}\sqrt{2}} + H.c.]$, $H_j= \frac{J_{\perp}}
{4\pi a}(X_{\uparrow\downarrow} F_{\downarrow}
^{\dagger}F_{\uparrow}+ H.c.)$. The effective Hamiltonian
can be written as

\begin{eqnarray}
H_{eff}^C = &&\sum_{k \sigma} E_k c_{k\sigma}^{\dagger}
c_{k\sigma} + {\sqrt{2}t}{\sqrt{2\pi a}}[X_{Ao}  c_{\uparrow}
c_{\downarrow} + H.c.]\nonumber\\
&& + (E_{d}^o -\frac{J_{\perp}}{4\pi a} )X_{AA} 
+ (E_d^o + \frac{J_{\perp}}{4\pi a} ) X_{BB} \nonumber\\
&&+\frac{\kappa_c}{2\pi \rho_o} (X_{AA} + X_{BB}-X_{oo})
(c^{\dagger}_{\uparrow}c_{\uparrow} + c^{\dagger}_{\downarrow} 
c_{\downarrow}) \nonumber\\
&&+\frac{\kappa_s}{2\pi \rho_o} (X_{AB} + X_{BA})
(c^{\dagger}_{\uparrow}c_{\uparrow} - c^{\dagger}_{\downarrow} 
c_{\downarrow}) 
\label{effhamC}
\end{eqnarray}
Here, $|A>= {1 \over \sqrt{2}} \sum_{\sigma}(-\sigma X_{\sigma o}
F_{\bar{\sigma}}^{\dagger})$ and $|B>= {1 \over \sqrt{2}} \sum_{\sigma}
(-X_{\sigma o} F_{\bar{\sigma}}^{\dagger})$.
Again, with the requirement on $\delta_0^{\sigma}$ such that 
the transformed potential scattering term vanishes, $\kappa_c$
and $\kappa_s$ can be written explicitly as 
$\kappa_c=\sqrt{2}(\delta_{\sigma}^{\sigma} +\delta_{\sigma}
^{\bar{\sigma}})$and $\kappa_s=\sqrt{2}(\delta_{\sigma}^{\sigma}
-\delta_{\sigma}^{\bar{\sigma}}-\pi)$, and
are non-zero only away from the Toulouse point.

In this effective Hamiltonian, the charge sector is described by a 
genuine \underline{charge Kondo} model. $|A>$ and $|o>$ play the
role of $|\uparrow>$ and $|\downarrow>$ of the spin Kondo problem
and should be thought of as objects carrying charge 2 and 0,
respectively. The transformed hybridization term is the direct
analog of the spin-flip term in the spin Kondo problem,
with a change of the charge quantum number by two replacing a
change of the spin quantum number by 1 in the latter. The
residual interaction in the charge sector, ${\kappa_c \over
2\pi \rho_o}$, is the analog of the longitudinal exchange
interaction in the spin Kondo problem, with the density
playing the role of the spin in the latter. In the conventional
notation, ${1 \over 2}J_{\perp}^{\rm charge} = 2t\sqrt{\pi a}$
and ${1 \over 4}J_{z}^{\rm charge}
= \frac{\kappa_c }{2\pi\rho_o}$. The essential difference
between the charge Kondo problem in this mixed valence
context and the spin Kondo problem lies in the
symmetry-breaking field. In the latter, the spin symmetry
guarantees that no explicit magnetic field term will be 
generated in the absence of an applied field. In our
charge Kondo problem, the particle-hole symmetry is
explicitly broken, and the symmetry-breaking field 
$h^{\rm charge} = E_{d}^o - \frac{J_{\perp}}{4\pi a}$
is in general non-zero. For the impurity problem, the condition
that the renormalized $h^{\rm charge}$ vanishes can be
achieved only through fine-tuning the bare $d-$level $E_{d}^o$
to a critical value $E_d^c$.

When $h^{\rm charge}=0$ is enforced, a zero temperature
quantum phase transition takes place as $\kappa_c$ is
increased through zero. The transition is characterized
by a Kosterlitz-Thouless transition in the charge
sector; the spin sector is not critical. The phenomenology
of the intermediate phase is recovered  on the negative
$\kappa_c$ side, to which the remaining of this section
is devoted. Here, the charge sector is described 
by the weak coupling fixed point of the charge-Kondo problem, 
while the spin excitations by the strong coupling,
Fermi liquid-like fixed point of the Kondo problem.
A spin-charge separation takes place. 

Within the charge sector, the impurity configuration in the ground 
state is entirely $|o>$ for $h^{\rm charge}<0$, and $|A>$ for 
$h^{\rm charge}>0$. This is the result of infinite charge 
susceptibility in the corresponding ferromagnetic charge Kondo
problem. Exactly at $h^{\rm charge}=0$, namely, when $E_d^0$
is tuned to the critical value $E_d^c= \frac{J_{\perp}}{4\pi a}$,
the impurity degrees of freedom in the ground state involve
an equal, incoherent, mixture of $|o>$ and $|A>$. Schematically,
the ground state wavefunction can be written as

\begin{eqnarray}
\phi = |A> \phi_{A} + |o> \phi_{o}
\label{gsmvtpc}
\end{eqnarray}
where $\phi_{A}$ and $\phi_{o}$ are the wave functions of the
conduction electrons such that $\phi$ solve a ferromagnetic 
Kondo model with zero magnetic field. 
With $h^{\rm charge}=0$, the intermediate mixed valence
dynamics applies at all temperatures. When $E_d^o$ is moved 
away from the critical value,  a finite cross-over
temperature $T_{co} \sim |E_d^0-E_d^c|$ emerges. The
intermediate mixed valence dynamics continue to apply
at $T > T_{co}$. At low temperatures ($T < T_{co}$),
however, the charge fluctuations become gapped out.
Such a cross-over can already be inferred from
the renormalization group trajectories in our previous 
work\cite{SiKotliar2}. 

The correlation functions that describe the critical dynamics 
associated with the intermediate mixed valence phase can be
calculated explicitly. Consider first the single particle 
Green function. Using $(d_{\sigma}^{\dagger})_{eff}=
e^{i\Phi_{s}/\sqrt{2}
\sigma}e^{-i\Phi_{c}/\sqrt{2}} (-{1 \over \sqrt{2}})(\sigma X_{Ao}
+X_{Bo})F_{\bar{\sigma}}$, we find that

\begin{eqnarray}
<{\rm T}_{\tau}d_{\sigma} (\tau) d^{\dagger}_{\sigma}(0)> && \sim
(\frac{\rho_o}{\tau})^{[{1 \over 2}+{1\over 2}(1-\sqrt{2}
{\kappa_c \over \pi})^2]}
\label{spintmed}
\end{eqnarray}
The fact that the exponent is interaction ($\kappa_c$)-dependent has
already been anticipated by the Coulomb gas
analysis\cite{SiKotliar1,SiKotliar2,KotliarSi}. What is new
here is the explicit demonstration that
the exponent of the single particle Green's function in the intermediate
phase is the sum of two terms, one from spin excitations and the
other from charge excitations. The spin contribution is
${1 \over 2}$, independent of interactions. The charge contribution,
on the other hand, is ${1\over 2}(1-\sqrt{2} {\kappa_c \over \pi})^2$ 
and varies as the interaction strength is changed. 
The exponent is always larger than one,
signaling a quasiparticle residue vanishing in a power-law
fashion in the intermediate phase -- a non-Fermi liquid indeed. 

For the spin-spin correlation functions, using $S^{+}_{eff}=
F_{\uparrow}^{\dagger}F_{\downarrow}e^{i\sqrt{2}\Phi_{s}}
\frac{1}{2} (X_{AA} - X_{BB} +X_{AB}- X_{BA} )$, and 
$S^{z}_{eff}=\frac{1}{2} (X_{AB} + X_{BA} )$, and following a
procedure similar to that used in calculating the spin-spin 
correlation functions in the second Toulouse point of 
the Kondo problem, we derive the following,

\begin{eqnarray}
<{\rm T}_{\tau}S^{-} (\tau) S^{+}(o)> &&\sim (\frac{\rho_o}{\tau})
^{2}\nonumber\\
<{\rm T}_{\tau}S^{z} (\tau) S^{z}(o)> &&\sim ({{\kappa_s} \over
2 \pi \rho_oh_s})^2 ({\rho_o \over \tau})^2
\label{spinintmed}
\end{eqnarray}
where $h_s={J_{\perp} \over 4\pi a}$. This establishes the Fermi-liquid
behavior of the spin-spin correlation functions.

We now turn to the density-density correlation function, using 
$\rho_{eff} = X_{AA} + X_{BB}-X_{oo}$. Due to the ergodicity 
breaking, there is a disconnected contribution to the density-density
correlation function that is $\tau -$independent. The connected
piece is algebraic, which we found to be,

\begin{eqnarray}
<{\rm T}_{\tau}\rho (\tau) \rho (o)>_{\rm connected}
&&\sim \frac{(\rho_ot)^2}{(-4\kappa_c)}
({\rho_o \over \tau})^{(-4\kappa_c)}
\label{densityintmed}
\end{eqnarray}
This algebraic piece, once again, has an interaction-dependent exponent.
We note that, the fact that the exponent vanishes with $\kappa_c$
has already been noted before in related problems\cite{Bhatt,Imbrie}.

Other correlation functions in the charge sector also have an
algebraic behavior with interaction-dependent exponents.
For instance, the excitonic correlation function, using
$(d_{\sigma}^{\dagger}c_{\sigma})_{eff}=e^{-i\sqrt{2}\Phi_{c}}
({1 \over \sqrt{2}})(X_{Ao}+\sigma X_{Bo})F_{\uparrow}F_{\downarrow}$,
has the following form,

\begin{eqnarray}
<{\rm T}_{\tau}(\sum_{\sigma}c_{\sigma}^{\dagger}d_{\sigma})(\tau)
(\sum_{\sigma}d_{\sigma}^{\dagger}c_{\sigma})(0)>
\sim (\frac{\rho_o}{\tau})^{2(1-{\kappa_c \over \sqrt{2}\pi})^2}
\label{excintmed}
\end{eqnarray}

Finally, we consider the pairing susceptibility. Following a similar 
procedure, we find that

\begin{eqnarray}
<{\rm T}_{\tau}(\sum_{\sigma}c_{\sigma}d_{\bar{\sigma}})(\tau)
(\sum_{\sigma}d_{\bar{\sigma}}^{\dagger}c_{\sigma}^{\dagger})(0)>
\sim (\frac{\rho_o}{\tau})^{({\kappa_c \over \sqrt{2}\pi})^2}
\label{pairingintmed}
\end{eqnarray}
which is enhanced compared to the Fermi-liquid case.
This makes it plausible that the ground state in the corresponding 
lattice model is superconducting. In that case, the intermediate
mixed valence dynamics would describe the physics in the normal state,
i.e., at temperatures between the transition temperature and some
upper cutoff energy scale.

To summarize, the intermediate phase has spin-charge separation, a
quasiparticle residue vanishing in a power-law fashion, self-similar
local correlation functions with interaction-dependent exponents.
These characteristics bear strong similarity to those of the
Luttinger liquid in one dimensional interaction
fermion systems\cite{Emery,Haldane81}. 

\section{A Point with Simple, but not Exactly Soluble, Effective
Hamiltonian}
\label{sec:toulD}

The three Toulouse points that we have discussed for
the mixed valence problem is exactly soluble. In this section,
we discuss one more point in the interaction parameter space
which is described by a simple effective Hamiltonian.
It arises from the choice of $\alpha =-(1-{1 \over \sqrt{2}})$ and
$\beta = \frac{1}{2\sqrt{2}}$. The Coulomb gas stiffnesses
are $\epsilon_t={1 \over 2}$ and $\epsilon_j = 2$.
In this case, there is a range of $\gamma$, $-{\sqrt{2} \over
4} \le \gamma \le {3 \over 4}\sqrt{2}-1$, that satisfies
the requirement that all the phase shifts $\delta_{\alpha}
^{\sigma}$ fall in the range $[-{\pi \over 2}, {\pi \over 2}]$:
$\delta_{\sigma}^{\sigma}=({3 \over 4}-{\sqrt{2} \over 2}
+ {\gamma \over \sqrt{2}})\pi$, $\delta_{\sigma}^{\bar{\sigma}}
=({\sqrt{2} \over 2}-{1 \over 4} + {\gamma \over \sqrt{2}})\pi$,
and $\delta_{o}^{\sigma}=(-{1 \over 4} + {\gamma \over
\sqrt{2}})\pi$. This range corresponds to a large
ferromagnetic $J_z$ and a large repulsive $V$ (and
attractive $V_p$). For instance, choosing $\gamma=\gamma_{min}
=-{\sqrt{2} \over 2}$ corresponds to $\delta(J_z/4)=-{{\sqrt{2}-1}
\over 2}\pi$, $\delta(V)={\pi \over 2}$, and $\delta(V_p) = -{\pi
\over 2}$, while chossing $\gamma=\gamma_{max}={3 \over 4}
\sqrt{2} -1 $ leads to $\delta(J_z/4)=-{\pi \over 2}$, 
$\delta(V)={\pi \over 2}$, and $\delta(V_p) = 
- {{\sqrt{2}-1} \over 2}\pi$.

Independent of the choice of $\gamma$,  $H_t=\frac{t}{\sqrt{2\pi a}} 
\sum_{\sigma}
(X_{\sigma o}F_{\sigma} e^{-i\Phi_{s}\sigma} + H.c )$,
$H_j= \frac{J_{\perp}}{4\pi a}  [X_{\uparrow\downarrow} 
F_{\downarrow}^{\dagger}F_{\uparrow}
e^{-i2\Phi_{s}} + H.c]$. As $H_j$ is strongly irrelevant, we can
take the $J_{\perp} = 0$ limit and consider $H_j$ as a
perturbation later. Introducing a fermion field $\eta=F_{\eta}
{1 \over \sqrt{2\pi a}}e^{-i\Phi_s}$, and $||\sigma>>
=|\sigma>F_{\sigma}F_{\eta}^{\dagger}$,
the effective Hamiltonian has the following form,

\begin{eqnarray}
H_{eff}^D= \sum_{k} E_{k} \eta^{+}_{k} \eta_{k} +
t [(X_{\uparrow   o} + X_{o \downarrow})\eta
+ H.c.] - E_{d}^o X_{oo}
\label{effD}
\end{eqnarray}

This Hamiltonian has the same form as that investigated analytically
and numerically in our earlier studies of a spinless two band model
in large dimensions\cite{SRKR}. It has also been found\cite{AER} to
arise within a particular rotation scheme in the bosonized form of 
the Anderson impurity model with additional screening channels.
What distinguishes this
effective Hamiltonian from those of the previous subsections
is the fact that, the operator
$(X_{\uparrow   o} + X_{o \downarrow})$ is now a rank-{\it 2} matrix.
In fact, $H_{eff}^D$ is a rank-2 generalization of the
Emery-Kivelson resonant-level model for the Toulouse
point of the two-channel spin$-{1 \over 2}$ Kondo 
problem\cite{EmeryKivelson}. The non-abelian nature of the phase
space makes the problem non-quadratic and not exactly soluble.
In the following, we establish the nature of the solutions in 
the two limits, $- E_d^o >> |t|, W$, and $ E_d^o >> |t|, W$.

Consider first the limit of $- E_d^o >> |t|, W$. Here, the
configurations $|\uparrow>$ and $\downarrow>$ lie at low energies,
while $|o>$ should be treated as a high energy configuration.
The hybridization $t$ term mixes the low energy and high energy
configurations, and can be eliminated with a canonical
transformation. This results in an effective Hamiltonian,
$H_o + (-{t^2 \over |E_d^o|}) (X_{\uparrow\uparrow}
-X_{\downarrow\downarrow}) (\eta^{\dagger}\eta -1/2)$. The sign of
the effective interaction makes the spin-flipping term even less
relevant. The result is a ground state with a double degeneracy.

For the opposite limit, $E_d^o >> |t|, W$, $|o>$ alone lies
at low energies, while $|\uparrow>$ and $\downarrow>$ are
high energy configurations that can be eliminated.
The resulting low energy behavior is a potential scattering problem.
The ground state is a singlet, and the low lying excitations have
the Fermi liquid form.
Therefore as the impurity level moves from far below to far above
the Fermi level, the system evolves from a non-Fermi liquid with 
a doublet ground state to an empty-orbital Fermi liquid.

We can in fact understand the doublet character of the
ground state in the limit of $- E_d^o >> |t|, W$ 
already from the renormalization group equations for the original
Hamiltonian, Eq. (\ref{hamil.us}).
The relevant renormalization group equations are given in
the Appendix B of Ref. (\cite{SiKotliar2}).
For a large negative $E_d^o$, the contribution to the scaling of
the spin-flip $J_{\perp}$ term from the second order in
hybridization $t$ term is small. The scaling of $J_{\perp}$ is
then entirely determined by $J_{z}$, as in the Kondo problem. 
Given that $J_z$ is ferromagnetic, the system remains a spin 
doublet at the fixed point. As $E_d^o$ moves closer to the Fermi
level, the contribution of the second order in $t$ term to the
scaling of $J_{\perp}$ becomes more important, and eventually
dominates the $J_z$ contribution.  These conclusions are
consistent with the  numerical calculations we have performed on 
the Hamiltonian (\ref{effD}).

Yet one more line of reasoning leads us to the same physical
picture of the phases of the  Hamiltonian (\ref{effD}).
We have shown earlier\cite{SRKR} that the Hamiltonian (\ref{effD})
maps onto the spinless resonant level model with particle-hole
symmetry and in the limit of vanishing $W/V_{\rm rl}$ and $W/t_{\rm rl}$
(where $W$, $V_{\rm rl}$ and $t_{\rm rl}$ are, respectively,
the conduction electron bandwidth, the interaction strength, and the
hybridization of the resonant level model). In this mapping, the
parameter t of Eq. (\ref{effD}) is of the order of $W$, while
$E_d^o = V_{rl}/2 - t_{rl}$. For a fixed $t_{rl}$, increasing 
$E_d^o$ in the Hamiltonian (\ref{effD}) is equivalent to 
increasing $V_{rl}$ in the resonant level model with particle-hole
symmetry. When $t_{rl}/W$ is small, the resonant level model
undergoes a phase transition from a doublet to
a singlet as $V_{rl}$ increases from a strongly attractive
value to a large repulsive one. The mixed valence regime
of the Hamiltonian (\ref{effD}) is the strong coupling
version of the critical regime associated with this phase
transition of the resonant level model. Whether the physics
of the resonant level model at large values of $t_{rl}/W$ and
$V_{rl}/W$ is smoothly connected to the physics of the
resonant level model at small $t_{rl}/W$ and $V_{rl}/W$, or 
a phase transition separates these two regimes, remain unclear
and is a problem deserving further investigation.

\bigskip
\section{Concluding Remarks}
\label{sec:conclu}

\subsection{Related Works}
\label{sec:comp}

The mixed valence problem is closely related to problems in the
macroscopic quantum tunneling (MQT) and macroscopic quantum coherence 
(MQC) where a few local degrees of freedom are coupled to a bath of
low energy excitations\cite{Leggett}. In the MQT and MQC literature, 
typically only two-level systems are considered. The mixed valence
problem amounts to a three-level generalization.

The mixed valence problem has historically been studied in the 
context of impurities embedded in a metal. It was stressed
early on that, in order to satisfy the Friedel
sum rule, screening interactions between
the impurity and additional channels of conduction electrons
have to be introduced\cite{Haldane.srn}. Similar screening
interactions have also been studied in the contexts of two-level
systems and macroscopic quantum tunneling\cite{Zawadowski,Hakim}.

In the context of high temperature superconductivity, it has been
suggested\cite{RuckensteinVarma} that the mixed valence regime of an
impurity model, with a repulsive impurity density-conduction electron
density interaction and a large number of screening conduction electron
orbitals, would be a local model exhibiting the novel non-Fermi liquid
characteristics of the marginal Fermi liquid, i. e.
logarithmically divergent charge and spin susceptibilities.
The numerical results of Ref. \cite{Perakis} indicate divergent
charge and spin susceptibilities near the mixed valence point.
However, the numerical results are not conclusive because the
susceptibility enhancement may be just due to a crossover to a local
moment regime as the impurity level is varied.
The divergence is especially difficult to see in the spin
susceptibility, as the susceptibility continues to increase when
$E_d^o$ is decreased through the transition regime\cite{Perakis}.

Within  the renormalization group analysis,
the effect of screening
channels is to modify the initial conditions of the renormalization
group flow\cite{SiKotliar1}. The additional screening channels
are passive
observers which slow down the response of the impurity degrees of
freedom (hence increase the orthogonality) but do not participate
in the formation of fixed points other than those within the
renormalization group classification. This is consistent
with the considerations of Giamarchi et. al.\cite{Giamarchi}
and Guinea et. al.\cite{Hakim}. From this perspective, the Anderson
model with additional screening channels has the same low energy
behavior as the generalized Anderson model considered here as
well as in our earlier works\cite{SiKotliar1,SiKotliar2,KotliarSi}. 

Recently, however, two groups\cite{Sire,Yulu} have sought to study
an exactly soluble point of the Anderson model with additional
screening channels. These works, using the bosonization method,
have reached conclusions that would signal a novel fixed point
unexpected from our previous renormalization group results.
One of the major conclusions in this work -- the arguments for which
being detailed in Secs. \ref{sec:sc1} and \ref{sec:sc2} -- is that
the fixed points discussed in Refs. \cite{Sire,Yulu} are the result
of using incorrect bosonization expressions for the fermion operators.
What was missing are the Klein
factors that keep track of the anticommutation relations of the
fermions of different spin species. When the correct bosonization
expressions are used, the results become compatible with our
previous renormalization group results.
We are therefore forced to the conclusion that, the only known
generic solutions to the spin$-{1 \over 2}$ mixed valence problem 
in the generalized Anderson impurity models are the three phases
identified within our previous renormalization group scheme.

\subsection{The Intermediate Phase in Realistic Models}
\label{sec:realistic}

The intermediate mixed valence state that we have identified represents
a new state of strongly correlated electron systems.
It will therefore be interesting to explore the phenomenological
consequences of this non-Fermi liquid state, and to search
for such a state in other strongly correlated
electron models. In this connection, two important issues
need to be addressed. 

First, within the generalized Anderson model that we have studied,
the interaction parameter regime for this intermediate phase corresponds
to a range of antiferromagnetic exchange interaction, and finite
attractive density-density interaction, between the impurity and
the conduction electrons [in addition to the large repulsive on-site
Hubbard interaction]. It is important to address how 
this attractive density-density interaction between the impurity and
conduction electrons can be generated from purely repulsive interactions
in more realistic impurity and lattice models. Several directions
have been explored along this direction. Additional screening channels
enhance the orthogonality effects\cite{RuckensteinVarma} and
effectively play the role of attractive density-density interactions.
Alternatively,
dynamical screening effects that arise from integrating out high energy
configurations in strongly correlated electron systems can generate
effective attractive density-density 
interactions\cite{KotliarSi}. From the perspective of a local approach
such as in the limit of infinite dimensions, this effect is especially
important since the effective bandwidth of the
self-consistent conduction electron bath is usually quite 
small\cite{KotliarSi,Moeller}. The generation of attractive
density-density interaction from purely repulsive interactions have
also been discussed in a class of multi-band models\cite{Aoki}.

Second, in general non-Fermi liquid states in impurity models can
be realized only through fine-tuning parameters. For the 
intermediate mixed valence phase, there is one parameter that 
needs to be fine-tuned, namely the impurity level $E_d^o$. This
places constraints on the realization of the intermediate state
in real materials displaying impurity physics. However, 
the situation becomes much better in real materials that are 
described by lattice models, as alluded to at the beginning of
this paper. In lattice models, the mixed valence condition
can be satisfied over a range of densities\cite{SKG,SiKotliar2}.

\subsection{\bf Summary}
\label{sec:sum}

In this paper, we have studied several exactly soluble points in the 
mixed valence regime of the generalized Anderson model. We found 
three such points and clarified the physics of each. 
Two of these points have the strong coupling, Fermi liquid 
behavior. The third Toulouse point describes the novel intermediate 
mixed valence phase.  Our explicit results about the correlation 
functions at this last Toulouse point clarified the excitation 
spectra in this new phase of strongly correlated electron systems.
Finally, we established that, once the anticommutation relations
between the bosonized forms of the fermion fields are taken
care of, the Toulouse-point results derived within the
bosonization formalism become consistent with our previous 
renormalization group results.

From a more general perspective, our results clearly illustrate 
the importance of the competition between {\it local} charge and spin
fluctuations in strongly correlated electron systems. In the 
intermediate phase, the competing low energy local charge and spin
fluctuations lead to a spin-charge separation, Fermi-liquid-like 
spin susceptibilities, non-Fermi-liquid charge susceptibilities,
non-Fermi liquid single particle spectral functions, and an
enhanced pairing susceptibility. This represents a new route 
towards spin-charge separation alternative to what 
leads to the Luttinger liquid in the interacting fermion
models in one dimension. In our case, the underlying physics is local
in nature, while in one dimension, it is dominated by long 
wavelength fluctuations. Nonetheless, the characteristics of the 
correlation functions in the intermediate phase have strong 
similarities to those of the Luttinger liquid.

\acknowledgments

We are grateful to E. Fradkin, T. Giamarchi, K. Ingersent,
A. J. Leggett, A. E. Ruckenstein, A. J. Schofield,
A. Sengupta, C. Sire, C. M. Varma, and YU Lu for useful
discussions. One of us (Q.S.) would like to acknowledge the
hospitality of the theory group at the AT\&T Bell Labs, where
part of this work was carried out. This work has been supported
by the NSF under the grant DMR 92-24000 (G.K.) and DMR 91-20000 (Q.S.).

\appendix{Bosonization for Impurity Problems}
\label{sec:bos} 

In this appendix, we summarize the bosonization procedure\cite{Emery}
relevant to our discussion. For the purpose of studying an 
impurity problem with contact interaction
at $\vec{r}=0$ only, the $S-$wave component of the conduction 
electrons alone needs to be kept. 
This $S-$wave component can in turn be written as a superposition of
an incoming component and an outgoing one defined on the 
half-axis $r \in [0, +\infty)$, with the boundary condition that 
these two components be equal at the origin\cite{AffleckLudwig}.
This boundary condition allows us to keep only one, say the 
incoming, component 
while simultaneously extending the problem to the full axis,
$x \in (-\infty, +\infty)$. Denoting 
this component as $\psi_{\sigma}(x)$, discarding the uncoupled 
higher-than-$S-$wave components of the conduction electrons,
and linearizing the energy dispersion about the Fermi level, 
we can rewrite the Hamiltonian for the non-interacting conduction 
electron bath, $H_o$, as follows,

\begin{eqnarray}
H_{o} = && \sum_{\sigma}{v_F } \int d x i\psi_{\sigma}
{{d \psi_{\sigma} }\over {d x}} 
\label{hopsi}
\end{eqnarray}
where $v_F$ is the Fermi velocity. The linearized 
conduction electron dispersion assumes the form
$E_{k} = v_{F} (k-k_F)$. The density of states
is $\rho(\epsilon)=\sum_k \delta (E_k-\epsilon)
= {1 \over {2 \pi v_F}}$.

We can now introduce a boson representation of the $\psi$-field. 
To keep track of the anticommutation relation between the 
fermion fields, we
write the boson representation of the fermion fields as follows,

\begin{eqnarray}
\psi_{\sigma}^{\dagger} (x) = F_{\sigma}^{\dagger} {1\over 
\sqrt{2\pi a}}{\rm e}^{i\Phi_{\sigma}(x)} 
{\rm e}^{ik_F x} 
\label{fermion}
\end{eqnarray}
Here $a$ is a cutoff scale which can be taken as a 
lattice spacing. $\Phi(x)$ is defined in terms of the Tomonaga bosons,

\begin{eqnarray}
\Phi_{\sigma}(x) = {2 \pi x \over L} N_{\sigma}
+ \sum_{q>0} \sqrt{2\pi \over q L } ( &&-i
b_{q\sigma}^\dagger e^{iqx-qa/2} \nonumber\\
&&+ i b_{q\sigma} e^{-iqx-qa/2})
\label{bos4}
\end{eqnarray}
In Eq. (\ref{bos4}), $b_{q\sigma}$ and $b_{q\sigma}^{\dagger}$
are the Tomonaga bosons,

\begin{eqnarray}
&&b_{q\sigma}^\dagger = \sqrt{2 \pi \over q L} \sum_k
\psi_{k+q~ \sigma}^\dagger \psi_{k\sigma}\nonumber\\
&&b_{q\sigma} = \sqrt{2 \pi \over q L} \sum_k \psi_{k\sigma}^\dagger
\psi_{k+q~\sigma}
\label{bos2.2}
\end{eqnarray}
Notice that $\Phi_{\sigma} (0)$ involves only the $q\ne0$ components of the
Tomonaga bosons. $L$ is the length of the dimension: 
$x \in [-L/2, L/2]$. $N_{\sigma}$ and $F_{\sigma}$ represent the zero
modes of the boson fields. $N_{\sigma}$ is the deviation of the
conduction electron occupation number from the ground state value. 
It represents the $q = 0$ counterpart of the finite $q$ Tomonaga boson
occupation number $n_{q\sigma} \equiv b^{\dagger}_{q\sigma}b_{q\sigma}$.
The boson Hilbert space is spanned by $|{N_{\sigma}}, {n_{q\sigma}}>$.
Within this Hilbert space, the operator $F_{\sigma}^{\dagger}$
raises $N_{\sigma}$ by one, while its adjoint, $F_{\sigma}$,
lowers $N_{\sigma}$ by one\cite{Haldane81,Heid,Neuberg}. These
are traditionally called the ``Klein factors''. They satisfy the
following relations,

\begin{eqnarray}
F_{\sigma}^{\dagger}F_{\sigma} = &&F_{\sigma} 
F_{\sigma}^{\dagger} = 1\nonumber\\
F_{\sigma}^{\dagger}F_{\bar{\sigma}} = &&-F_{\bar{\sigma}} 
F_{\sigma}^{\dagger}\nonumber\\
F_{\sigma}F_{\bar{\sigma}} = &&-F_{\bar{\sigma}} F_{\sigma}
\label{raising}
\end{eqnarray}
They commute with $b_{q\sigma}$ and $b_{q\sigma}^{\dagger}$,
for $q \ne 0$.

In this boson representation, the non-interacting Hamiltonian Eq.
(\ref{hopsi}) has the following form,

\begin{eqnarray}
H_o = {v_F \over {4\pi}} \sum_{\sigma} \int_{-\infty}^{+\infty}
dx ({{d\Phi_{\sigma} (x)} \over {dx}})^2
\label{bosh0}
\end{eqnarray}
and the density operator
\begin{eqnarray}
\rho_{\sigma} (x) \equiv \psi_{\sigma}^{\dagger} \psi_{\sigma}
= {1 \over {2 \pi}} {d \Phi_{\sigma} \over d x}
\label{bos5}
\end{eqnarray}

\bigskip
\appendix{Atomic Expansion}
\label{sec:atomic} 

Consider first the Toulouse point II of the Kondo problem
discussed in Sec. \ref{sec:kondo-toulII}. Given that
the vicinity of this Toulouse point corresponds to
$J_{z} >> J_{\perp}, W$, we can carry out an expansion
in terms of ${J_{\perp} \over J_z}$ and ${W \over J_z}$.
For this purpose, 
we rewrite the original Kondo Hamiltonian as

\begin{eqnarray}
H = &&W \sum_{i=0,\sigma}^N( c^{\dagger}_{i\sigma} c_{i+1\sigma } 
+ H.c.) + {J_{\perp} \over 2}( S_{+} c^{\dagger}_{\downarrow}
c_{\uparrow} + S_{-}c^{\dagger}_{\uparrow}c_{\downarrow} ) \nonumber\\
&&+ { J_{z} \over 4} S_{z} \sum_{\sigma} \sigma 
c^{\dagger}_{0\sigma} c_{0\sigma}
\end{eqnarray}
where  $c^{\dagger}_{i\sigma} $ creates a Wannier orbital 
at site $i$ in the Wilson basis\cite{Wilson}. To the
leading order, we need simply to diagonalize an atomic
problem defined in the Hilbert space of the impurity spin
doublet and $c^{\dagger}_{0\sigma} $. The lowest 
energy atomic states are given 

\begin{eqnarray}
|1> = &&|\uparrow>_d |\downarrow>_0 \nonumber\\
|2> = &&|\downarrow>_d |\uparrow>_0
\label{atomickondo.r}
\end{eqnarray}
For finite $J_{\perp}$ and $W$, these low energy atomic
states become coupled with the high energy atomic states,
$|A\sigma> =|\sigma>_d|2>_0$, $|B\sigma > =|\sigma>_d|0>_0$, 
and $|C\sigma >=|\sigma>_d|\sigma>_0$. Integrating out these
high energy states via a canonical transformation

\begin{eqnarray}
S={W \over J_z} (&&-X_{A\uparrow,1}c_{1\uparrow} + 
X_{B\uparrow,1}c_{1\downarrow}^{\dagger} \nonumber\\
&&+X_{A\downarrow,2}c_{1\downarrow} + 
X_{B\downarrow,2}c_{1\uparrow}^{\dagger} -H.c.) \nonumber
\label{canonkondoatomicapen}
\end{eqnarray}
leads to an effective Hamiltonian $H_{eff} = e^{S} H e^{-S}$.
To the second order in ${W/J_z}$,

\begin{eqnarray}
H_{eff}^{2\prime}= &&H_o^{\prime}+J_{\perp} (X_{12} + X_{21})
\nonumber\\
&&+ J_z^{\prime}(X_{11} - X_{22}) \sum_{\sigma} \sigma 
c^{\dagger}_{1\sigma} c_{1\sigma}
\label{effkondoatomicapen}
\end{eqnarray}
where
\begin{eqnarray}
H_o^{\prime} = W \sum^{N}_{i=1,\sigma}
(c^{\dagger}_{i\sigma} c_{i+1\sigma} + H.c.)
\label{ho'}
\end{eqnarray}
and $J_z^{\prime} \sim \frac{W^{2}}{J_z}$.

The effective spin operators are
, $S^{z}_{eff} =
$ and
$S^{+}_{eff}=
$ are given by

\begin{eqnarray}
S^{z}_{eff} &&\equiv e^{S} S^{z} e^{-S}
=  {1 \over 2} (X_{11} - X_{22}) \nonumber \\
S^{+}_{eff} &&\equiv e^{S} S^+ e^{-S}
=  2({W \over J_z})^2 c^{\dagger}_{1\uparrow} c_{1\downarrow}
X_{12}  
\label{szxykondoatomicappen}
\end{eqnarray}
Eqs. (\ref{effkondoatomicapen}-\ref{szxykondoatomicappen}) are the
direct analogs of Eqs. (\ref{spinboson}) and 
(\ref{szxykondotoul2}).

We now turn to the strong coupling Toulouse point I of the mixed valence
problem discussed in Sec. IV. The low energy atomic states are,

\begin{eqnarray}
|1> = &&|\uparrow>_d |\downarrow>_0 \nonumber\\
|2> = &&|\downarrow>_d |\uparrow>_0\nonumber\\
|O> = &&|0>_d|2>_0
\label{atomicmv1}
\end{eqnarray}
The mixed valence condition amounts to the requirement on $E_d^o$
such that the energy difference between $|1>$, $|2>$ and $|O>$,
$2E_d^{\prime}=E_d^o+V+|V_p|-{J_z \over 4}$, is small.
The canonically transformed Hamiltonian is

\begin{eqnarray}
H_{eff}^{A\prime} = &&H_o^{\prime} + E_d^{\prime}(X_{11}+X_{22}-
X_{OO}) +J_{\perp} (X_{12}+X_{21})\nonumber\\
&& + t(X_{1O}-X_{2O}+H.c.)\nonumber\\
&&+J_z^{\prime}(X_{11}-X_{22})\sum_{\sigma}\sigma
c_{1\sigma}^{\dagger} c_{1\sigma}\nonumber\\
&&+V_z^{\prime} (X_{11}+X_{22})\sum_{\sigma}c_{1\sigma}^{\dagger} 
c_{1\sigma}
\label{effHmv1}
\end{eqnarray}
where $V_z^{\prime}=\frac{W^{2}}{V}$. The canonically-transformed
operators are
\begin{eqnarray}
(d_{\uparrow}^{\dagger})_{eff} \sim && ({W \over J_z}+{W\over V})
c_{1\uparrow}^{\dagger}X_{1 O}\nonumber \\
(d_{\downarrow}^{\dagger})_{eff} \sim && -({W \over J_z}+{W\over V})
c_{1\downarrow}^{\dagger}X_{2 O}\nonumber \\
S^{+}_{eff} \sim &&  2({W \over J_z})^2 
c^{\dagger}_{1\uparrow} c_{1\downarrow} X_{12}  \nonumber\\
S^{z}_{eff} \sim && {1 \over 2} (X_{11} - X_{22}) \nonumber\\
\rho_{eff} \sim && (X_{11} + X_{22}-X_{OO})
\label{szxymv1atomic}
\end{eqnarray}
These expressions are the direct analogs of the expressions
derived using the bosonization representation,
Eqs. (\ref{effha}) and (\ref{transopmv1}).

Finally, we consider the intermediate Toulouse point of Sec. VI.
The low energy atomic states are,

\begin{eqnarray}
|A> = &&u {1 \over \sqrt{2}} \sum_{\sigma} \sigma 
|\sigma>_d|\bar{\sigma}>_0 + v |0>_d|2>_0 \nonumber\\
|B> = && {1 \over \sqrt{2}} \sum_{\sigma} |\sigma>_d
|\bar{\sigma}>_0\nonumber\\
|O> = &&|0>_d|0>_0
\label{atomicmv3}
\end{eqnarray}
The canonically transformed Hamiltonian is

\begin{eqnarray}
H_{eff}^{C\prime} = &&H_o^{\prime} + J_{\perp} [-X_{AA}+X_{BB}]
+t' (X_{AO} c_{1\uparrow}c_{1\downarrow} +H.c)\nonumber\\
&&+V' (X_{AA}-X_{OO})(\sum_{\sigma}c_{1\sigma}^{\dagger}c_{1\sigma})
\label{effHmv3}
\end{eqnarray}
where $t' \sim -{{2vW^2} \over V}$ and $V' \sim 
 -{{(1+v^2)W^2} \over {2V}}$.  The canonically transformed 
$S_{eff}^{+}$, $S_{eff}^z$, and $\rho_{eff}$ are similar to
those given in Eq. (\ref{szxymv1atomic}), while

\begin{eqnarray}
(d_{\sigma}^{\dagger})_{eff} \sim && -{1 \over \sqrt{2}}
({W \over J_z}+{W\over |V|}) c_{1\bar{\sigma}}
(\sigma X_{AO} + X_{BO})
\label{dmv3atomic}
\end{eqnarray}


\begin{references}
 
\bibitem[*]{rice}
On leave from Physics Department, Rice University, Houston,
TX 77251-1892

\bibitem{AndersonYuval} P. W. Anderson, G. Yuval and D.R. Hamman, 
Phys. Rev. B1, 4464 (1970).

\bibitem{Wilson} K. G. Wilson, Rev. Mod. Phys. 47, 773 (1975).

\bibitem{Nozieres} P. Nozieres, J. Low Temp. Phys. 17, 31 (1974).

\bibitem{AndreiWiegmann} N. Andrei, K. Furuya, and J. H. Lowenstein, 
Rev. Mod. Phys. 55, 331 (1983); A. M. Tsvelick and P. B. Wiegmann,
Adv. Phys. 32, 453 (1983).

\bibitem{VarmaYafet} C. M. Varma and Y. Yafet, Phys. Rev. B13, 
2950 (1976); C. M. Varma, Rev. Mod. Phys. 48, 219 (1976).

\bibitem{Haldane} F.D.M. Haldane, Phys. Rev. Lett. 40,416(1978);
J Phys. C11, 5015 (1978).

\bibitem{Krishnamurthy} H. R. Krishnamurthy,
K. G. Wilson and J. W. Wilkins, Phys. Rev. B21, 1044 (1980).

\bibitem{Bickers} For a review see N. E. Bickers, Rev. Mod. Phys.
59, 845 (1987).

\bibitem{SiKotliar1} Q. Si and G. Kotliar, Phys. Rev. 
Lett. 70, 3143 (1993).

\bibitem{SiKotliar2} Q. Si and G. Kotliar, Phys. Rev. B48, 13881 (1993).

\bibitem{KotliarSi} G. Kotliar and Q. Si, Phys. Scrip. T49, 165 (1993).

\bibitem{MorelAnderson} P. Morel and P. W. Anderson, Phys. Rev. 125, 
1263 (1962).

\bibitem{KaneFisher} C. L. Kane and M. P. A. Fisher,
Phys. Rev. B46, 15233 (1992).

\bibitem{Furusaki} A. Furusaki and N. Nagaosa, Phys. Rev. B47,
4631 (1993).

\bibitem{Bethe} While the Kondo problem and the usual Anderson model
with infinite bandwidth are exactly soluble using the Bethe 
ansatz method\cite{AndreiWiegmann} and conformal field
theory\cite{AffleckLudwig}, the Anderson model with an 
additional density-density interaction has not been solved this way.

\bibitem{Toulouse} G. Toulouse, C. R. Acad. Sci. 268, 1200 (1969).

\bibitem{Perakis} I. Perakis, C. Varma, and A. Ruckenstein,
Phys. Rev. Lett. 70, 3467(1993).

\bibitem{Sire} C. Sire, C. M. Varma, A. E. Ruckenstein, and T. Giamarchi,
Phys. Rev. Lett. 72, 2478 (1994).

\bibitem{Yulu} G. M. Zhang and Lu Yu, Phys. Rev. Lett. 72, 2474;
G. M. Zhang, Z. B. Su, and Lu Yu, preprint (1995).

\bibitem{EmeryKivelson} V. J. Emery and S. A. Kivelson,
Phys. Rev. B46, 10812 (1992).

\bibitem{Finkelstein} P.B. Wiegmann and A.M. Finkelstein, Sov. Phys. JETP,
48, 102 (1978); P. Schlottmann, Phys. Rev. B25, 4815 (1982).

\bibitem{Blume} M. Blume, V. J. Emery, and A. Luther, 
Phys. Rev. Lett. 26, 1547 (1971); V. J. Emery and A. Luther,
Phys. Rev. B9, 215 (1974).

\bibitem{Leggett} For a recent review see A. J. Leggett, 
S. Chakravarty, A. T. Dorsey, M.P.A. Fisher, 
and W. Zwerger, Rev. Mod. Phys. 59, 1 (1987);
{\it ibid.} 67, 215 (Erratum) (1995).

\bibitem{AffleckLudwig} I. Affleck and A. W.W. Ludwig, Nucl. Phys.
B360,  641 (1991).

\bibitem{Emery} V. J. Emery, in {\it Highly Conducting
One-dimensional Solids}, Eds. J. T. Devreese {\it et al.} (Plenum, New
York, 1979); J. Solyom, Adv. Phys. 28, 201 (1979).

\bibitem{Haldane81}F.D.M. Haldane, J. Phys. C14, 2585 (1981).

\bibitem{Heid} R. Heidenreich, R. Seiler, and 
D. A. Uhlenbrock, J. Stat. Phys. 22, 27 (1980) and references
therein.

\bibitem{Neuberg} H. Neuberger, Tel Aviv University Thesis (1975).

\bibitem{finiteT} At finite temperature $T={1 \over \beta}$,
$1/\tau$ is replaced by ${\pi \over \beta} / sin ({\pi \over
\beta}\tau )$.

\bibitem{Hakim} F. Guinea, V. Hakim, and A. Muramatsu, Phys. Rev.
B32, 4410 (1985).

\bibitem{Spohn} H. Spohn, Comm. Math. Phys. 123, 227 (1989) and references
therein.

\bibitem{Weiss} M. Sassetti and U. Weiss, Phys. Rev. Lett. 65, 2262 (1990).

\bibitem{Chakravarty} S. Chakravarty and J. Rudnick, 
Phys. Rev. Lett. 75, 501 (1995).

\bibitem{Bhatt}J. Bhattacharjee, S. Chakravarty, J. L. Richardson,
and D. J. Scalapino, Phys. Rev. B24, 3862 (1981).

\bibitem{Imbrie} J. Z. Imbrie and C. M. Newman, Commun. Math. Phys.
118, 303 (1988).

\bibitem{SRKR} Q. Si, M. Rozenberg, G. Kotliar, and A. Ruckenstein,
Phys. Rev. Lett. 72, 2761 (1994).

\bibitem{AER} A. E. Ruckenstein, unpublished.

\bibitem{Haldane.srn} F. D. M. Haldane, Phys. Rev. B15, 2477 (1977).

\bibitem{Zawadowski} K. Vladar, G. T. Zimanyi, and A.
Zawadowski, Phys. Rev. Lett. 56, 286 (1986), and references therein.

\bibitem{RuckensteinVarma} A. E. Ruckenstein and C. M. Varma, 
Physica C185-189, 134 (1991).

\bibitem{Giamarchi} T. Giamarchi, C. Varma, A. Ruckenstein, and P.
Nozieres, Phys. Rev. Lett. 70, 3967 (1993).

\bibitem{Moeller} G. Moeller, Q. Si, G. Kotliar, M. Rozenberg, and 
D. S. Fisher, Phys. Rev. Lett. 74, 2082 (1995).

\bibitem{Aoki} K. Kuroki and H. Aoki, Journ. Supercond. 7,
577 (1994).

\bibitem{SKG} Q. Si, G. Kotliar, and A. Georges, Phys. Rev. B46,
1261 (1992).

\end{references}
\end{document}